\documentclass[fleqn,10pt]{wlscirep}

\usepackage{lipsum}
\usepackage[utf8]{inputenc}
\usepackage[T1]{fontenc}
\usepackage{setspace}
\usepackage{amsmath}
\usepackage{graphicx}
\usepackage{float}
\usepackage{microtype}
\usepackage{amssymb}
\usepackage{amsthm}
\usepackage{physics}

\usepackage{tikz}
\usepackage{hyperref}
\usepackage{pgf-pie}
\usepackage{graphicx}
\usepackage{pdfpages}
\usepackage{color}
\usepackage{subcaption}
\usepackage[]{algorithm2e}
\usepackage{bbold}
\usepackage{mathtools}
\usepackage{enumerate}
\usepackage{tcolorbox}
\usepackage{adjustbox}
\usepackage{floatflt}
\usepackage{comment}
\usepackage{svg}
\usepackage[OT1]{fontenc}
\usepackage{hyperref}

\usepackage{ulem}

\newcommand{\M}{\mathcal{M}}
\newcommand{\sumkl}{\sum_{k,l = 0}^{d-1}}

\newcommand{\Pkl}{P_{k,l}}
\newcommand{\Okl}{\Omega_{k,l}}

\newcommand{\E}{\mathcal{E}}
\newcommand{\K}{\mathcal{K}}
\newcommand{\ckl}{c_{k,l}}

\title{Comparing Bound Entanglement of Bell Diagonal Pairs of Qutrits and
Ququarts}

\author[1,*]{Christopher Popp}
\author[1,**]{Beatrix C. Hiesmayr}

\affil[1]{University of Vienna, Faculty of Physics, Währingerstrasse 17, 1090 Vienna, Austria}
\affil[*]{christopher.popp@univie.ac.at}
\affil[**]{Beatrix.Hiesmayr@univie.ac.at}

\begin{abstract}
  We compare the classification as entangled or separable of Bell diagonal
  bipartite qudits with positive partial transposition (PPT) and their
  properties for different dimensions. For dimension $d \geq 3$, a form of
  entanglement exists that is hard to detect and called bound entanglement due
  to the fact that such entangled states cannot be used for entanglement
  distillation. Up to this date, no efficient solution is known to differentiate
  bound entangled from separable states. We address and compare this problem
  named separability problem for a family of bipartite Bell diagonal qudits with
  special algebraic and geometric structures and applications in quantum
  information processing tasks in different dimensions. Extending analytical and
  numerical methods and results for Bell diagonal qutrits ($d=3$), we
  successfully classify more than $75\%$ of representative Bell diagonal PPT
  states for $d=4$. Via those representative states we are able to estimate the
  volumes of separable and bound entangled states among PPT ququarts ($d=4$). We
  find that at least $75.7\%$ of all PPT states are separable, $1.7\%$ bound
  entangled and for $22.6\%$ it remains unclear whether they are separable or
  bound entangled. Comparing the structure of bound entangled states and their
  detectors, we find considerable differences in the detection capabilities for
  different dimensions and relate those to differences of the Euclidean geometry
  for qutrits ($d=3$) and ququarts ($d=4$). Finally, using a detailed visual
  analysis of the set of separable and bound entangled Bell diagonal
  states in both dimensions, qualitative observations are made that
  allow to better distinguish bound entangled from separable states.
\end{abstract}

\begin{document}
\flushbottom
\maketitle
\thispagestyle{empty}
\newpage

\section*{Introduction}
Quantum technology leverages quantum phenomena for better performance than
classical methods for applications like computing, communication, simulation,
metrology and cryptography \cite{distQuantComp, QuantCryptBell, quantSimReview,
  twoStepQuantDirCom, oneStepQSDC}. Quantum information theory provides the
theoretical formalism for processing tasks using quantum mechanical systems
\cite{nielsen}. One of the characteristic properties of a quantum system that
allows realizing information processing with superior performance compared to
classical systems is entanglement. Besides its relevance for our general
understanding of nature and the interpretation of quantum theory \cite{epr,
  bellEpr,loopholeFreeBell} it provides one of the main resources to realize
applications in various fields ranging from quantum teleportation to medical
applications for the detection of cancer cells \cite{Moskal_2016, jpet,
  3photonDec, comptonMUB}. The simplest system to observe entanglement is the
bipartite system of two two-level quantum systems, called qubits. Currently,
most applications are based on these quantum systems of dimension $d=2$, but
recently, interest in higher dimensional systems like ``qutrits'' for $d=3$,
``ququarts'' for $d=4$ or ``qudits'' for general $d$ is growing due to potential
advantages and new observable phenomena \cite{highDimQCReview,
  qditsQCWang}. Bell states \cite{bellStates} are special sets of entangled
states which can be used as basis for the corresponding Hilbert space. They are
highly relevant for applications due to the fact that they are maximally
entangled states. Originally introduced for $d=2$, they can be
generalized for higher dimensions \cite{Sych, teleportingWeyl, baumgartner1}. \\
In this paper we analyze mixtures of maximally entangled bipartite Bell states,
with focus on $d=3$ and $d=4$. Those states are locally maximally mixed, meaning
that there is no correlation in the respective subsystems. Depending on the
mixing probabilities of the $d^2$ pure Bell basis states, a general mixed state
can be entangled or not, in which case it is separable. The Peres-Horodecki
criterion, also known as PPT (positive partial transposition) criterion
\cite{peres,horodeckiCriterion}, provides an efficient method to detect a state
as entangled, if the partially transposed density matrix of a given quantum
state has at least one negative eigenvalue, in which case the state is called
``NPT''. Otherwise it is called ``PPT''. For $d=2$, all entangled states are
NPT, but for $d \geq 3$, also PPT entangled states exist
\cite{distillation}. While NPT entangled states can be ``distilled''
\cite{distillation} to result in fewer strongly entangled states, this process
is not possible for PPT entangled ones. For this reason PPT entanglement is also
called ``bound'' entanglement, which has been extensively investigated since its
discovery by the Horodecki family, e.g. in Refs.~\cite{Halder_BE, lockhart_e,
  Bruss_BE, Slater2019JaggedIO,choi, Chru_EW, BaeLinkingED,
  Bae_structPhysApprox, Korbicz2008StructuralAT, Huber_hdent,
  augus_optEw}. 2014, it has also been observed in experiment, using photons
entangled in their orbital angular momentum \cite{hiesmayrLoeffler}. Many
applications like teleportation or superdense coding \cite{nielsen} require
strongly entangled states for reliable performance. However, if a given state is
transformed to a bound entangled one, this resource is bound for immediate
application, since it cannot be used to distill strongly entangled states. For
this reason, it is important to know about structure of bound entanglement in a
given system and to be able to detect those states reliably, so that operations
that result in binding the resource entanglement in certain states can be
avoided. However, the ``separability problem'' to differentiate separable and
PPT entangled states has been proved to be NP-hard \cite{nphard, nphard-strong}
in general and lacks an efficient solution if the dimension of the system is not
small. Existing methods to detect PPT entangled states \cite{choi, Chru_EW,
  BaeLinkingED, Bae_structPhysApprox, Korbicz2008StructuralAT, Huber_hdent,
  augus_optEw} are often strongly limited in the number of states they can
detect and are not efficient in higher dimensions.  Likewise, no efficient
method to decide whether a PPT state is separable or bound entangled states is
known for Bell diagonal qudits, which are known to be highly relevant for
practical applications \cite{Werner_2001}. However, special families of these
states have strong symmetries that can be leveraged for the analytical and
numerical analysis of its properties regarding the entanglement structure
\cite{baumgartner1, hiesmayrLoeffler,baumgartnerHiesmayr}. In particular, the
analytical structure of mixed Bell states generated by Weyl-Heisenberg
transformations \cite{teleportingWeyl} allows to derive several criteria to
detect separability and entanglement \cite{baumgartner1,
  baumgartnerHiesmayr}. Furthermore, an efficient geometric representation of
the states, symmetries and entanglement
witnesses \cite{EWs, EW20} makes the system well applicable for numerical methods. \\
Recently, analytical and numerical methods were combined to solve the
separability problem for the system in three dimensions in an ``almost
complete'' way \cite{PoppACS}. Given any unknown PPT state, the developed
methods allow the classification of this state as separable or bound entangled
with a probability of success of $95 \%$. Moreover the classification allows the
determination of the relative volumes of separable, PPT and NPT entangled
states. It was further shown, that a significant share of the PPT states of
bipartite, Bell diagonal qutrits are bound entangled (at least $13.9\%$), making
this system exceptionally well suited to study this exotic form of entanglement
regarding its detection, use in information processing tasks and implications
for nature. It is expected that the dimension of the system has a large
influence on the structure and the relative shares of entanglement classes,
which is a focus of this contribution. While approximations of the relative
volumes of separable states in general systems in dependence on the dimension
exist \cite{volOfSep1, volOfSep2}, the precise
numbers depend on the specific system and are not known. \\
The aim of this work is to extend and apply those methods, used to successfully
characterize the system for qutrits \cite{PoppACS}, for $d=4$, to draw
conclusions about the structure of entangled and separable states as well as the
effectiveness of their detectors and to compare the results to $d=2$ and
$d=3$. The paper is organized as follows: First, the system to be analyzed is
defined and relevant methods to generate states and to investigate its
entanglement structure are presented for general dimension. Second, we analyze
the set of PPT states for $d=4$. We quantify the share of this set in the total
system and the relative volumes of separable and (bound) entangled states within
and compare to other dimensions. Then, the applied criteria to detect
separability and entanglement are compared for their effectiveness in different
dimensions. Finally, we leverage the special properties of the system to
visualize the set of separable and bound entangled states for $d=3$ and
$d=4$. The visual analysis demonstrates relations between the algebraic
structure of Bell diagonal mixtures and geometric restrictions on the set of
separable and bound entangled states. This might be leveraged for the detection of
bound entanglement and separability.
\newpage

\section*{Methods}
\begin{figure}[H]
  \centering
  \begin{minipage}{.5\textwidth}
    \centering
    \includegraphics[width=\linewidth]{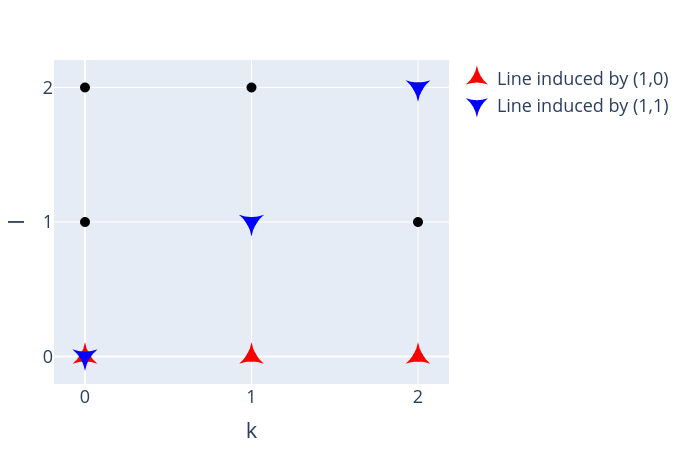}
    \captionof{figure}{Phase space and exemplary induced \\subgroups called
      ``lines'' for $d=3$.\\ Figure created with Ref.\cite{plotlyJs}.} 
    \label{phaseSpace3}
  \end{minipage}%
  \begin{minipage}{.5\textwidth}
    \centering
    \includegraphics[width=\linewidth]{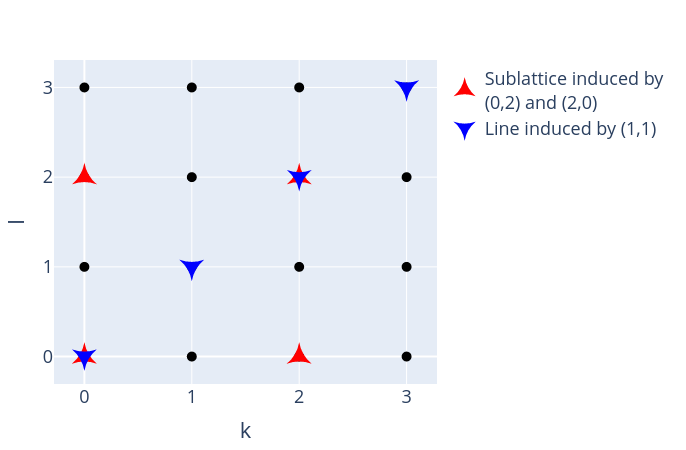}
    \captionof{figure}{Phase space and exemplary induced \\subgroups called
      ``lines'' and ``sublattices'' for $d=4$.\\ Figure created with Ref.\cite{plotlyJs}.}
    \label{phaseSpace4}
  \end{minipage}
\end{figure}

Consider the Hilbert space $\mathcal{H} = \mathcal{H}_1 \otimes \mathcal{H}_2$
for the bipartite system of two qudits of dimension $d$. In this work we
analyze mixtures of maximally entangled orthonormal Bell states
$\ket{\Okl} \in \mathcal{H}$ with $k, l = 0, 1, \cdots , (d-1)$ generated by
applying the Weyl operators \cite{teleportingWeyl} $W_{k,l}$ to one qudit of the
shared maximally entangled state
$ \ket{\Omega_{00}} \equiv \frac{1}{\sqrt{d}} \sum_{i = 0}^{d-1} \ket{ii}$:
\begin{gather}
  \label{bellStates}
  \ket{\Okl} \equiv W_{k,l} \otimes \mathbb{1}_d \ket{\Omega_{00}}
\end{gather}
where
$ W_{k,l} \equiv \sum_{j=0}^{d-1}w^{j \cdot k} \ket{j} \bra{j+l \pmod d},~w =
e^{\frac{2 \pi i}{d}}$. Mixing the density matrices, or ``Bell projectors'',
$P_{kl} \equiv \ket{\Okl}\bra{\Okl}$ with mixing probability $c_{k,l}$ defines
Bell diagonal states with respect to the above defined Weyl operators and the
system of interest for this work:
 \begin{gather}
\M_d \equiv \lbrace \rho = \sumkl c_{k,l} P_{k,l}~ |~
\sumkl c_{k,l} = 1, c_{k,l} \geq 0  \rbrace
\end{gather}
By taking the partial trace with respect to one of the subsystems, the reduced
state of any state in $\M_d$ is maximally mixed, so all information is in the
correlation of the combined state and not in the subsystems themselves. States
with this property are called ``locally maximally mixed''. Any state of $\M_d$
is equivalent to a point in $d^2$ dimensional Euclidean space by identifying
the mixing probabilities $c_{k,l}$ with coordinates in real space. Due to the
normalization of the $c_{k,l}$, the set of these points forms a
standard simplex. Referring to the ``magic Bell basis'' of Wootters and Hill
\cite{magicBellBasis} $\M_d$ is also known as ``magic simplex''
\cite{baumgartner1,baumgartnerHiesmayr,baumgartnerHiesmayr2}. \\
The properties of the Weyl operators $W_{k,l}$ imply a linear ring structure or
``discrete phase space'' for operators indexed by the tuples $(k,l)$ based on
succeeding application of these operators \cite{baumgartner1}. This can be seen
via the Weyl relations \cite{teleportingWeyl} (addition defined modulo $d$):
\begin{gather}
  W_{k_1,l_1}W_{k_2,l_2} = w^{l_1 k_2}~W_{k_1+k_2, l_1+l_2}  \\
  W_{k,l}^\dagger = w^{k l}~W_{-k, -l} = W_{k,l}^{-1}
\end{gather}
In Fig.\ref{phaseSpace3} and Fig.\ref{phaseSpace4} we visualize this phase space
as lattice of $d \times d$ vertices, each vertex corresponding to the Weyl
operator (and thus also to a Bell state via eq.(\ref{bellStates})) with
according indices $(k,l)$. Depending on the dimension $d$, several subgroups
exist and for this work subgroups containing $d$ elements are of special
relevancy, as they can be related to the structure of the sets of separable and
bound entangled states. In general, a subgroup is defined by its generating
elements. In case $d$ is prime, all subgroups of $d$ elements are generated by
one of the Weyl operators. Highlighting the vertices (here red and blue)
corresponding to a subgroup generated by $W_{k,l}$ induces ``lines'' in the
discrete phase space (see Fig.\ref{phaseSpace3}). This is different for
non-prime dimensions, where subgroups of $d$ elements can additionally be
generated by two Weyl operators whose indices contain proper divisors of $d$ in
which case ``sublattices'' are formed in the phase space (see
e.g. Fig.\ref{phaseSpace4}). For more details, consult
Ref.\cite{baumgartner1}. We can also use Fig.\ref{phaseSpace3} and
  Fig.\ref{phaseSpace4} to specify ``subgroup states'' by assigning each
  highlighted subgroup to a state which consists of equal mixtures of all
  corresponding Bell states.
\ \\

\subsection*{Relevant subsets}
\label{relSub}
The properties of $\M_d$ allow the definition of special subsets related to the
entanglement properties of contained states, which have been investigated with
respect to separability and (bound)
entanglement. \cite{baumgartnerHiesmayr2,BaeQuasipure}
\\ \ \\
\textbf{Enclosure polytope} \\
The enclosure polytope is a superset of all states with positive partial
transposition. It was shown \cite{baumgartner1} that all states that have at
least one mixing probability $c_{k,l}$ exceeding $1/d$ are necessarily entangled and can
be detected by the Peres-Horodecki criterion (PPT criterion). As NPT entangled
states they can be distilled by local operation and classical communication
(LOCC), and are therefore called ``free'' and not ``bound'' entangled
\cite{distillation}. The enclosure polytope is defined as:
\begin{gather}
  \label{encP}
\E_d \equiv \lbrace \rho = \sumkl c_{k,l} P_{k,l}~ |~
\sumkl c_{k,l} = 1, c_{k,l} \in [0, \frac{1}{d}]  \rbrace
\end{gather}
Using the representation of $\M_d$ in Euclidean space, $\E_d$ forms a bounded polytope.
\\
\newline
\textbf{Kernel polytope} \\
For each of the subgroups of $d$ elements induced by the Weyl operators (indexed
by $\alpha$), a special ``subgroup'' or ``sublattice state'' $\rho_{\alpha}$ can
be defined, which is known to be a separable state \cite{baumgartner1}. In
general, these subgroup states are equal mixtures of $d$ Bell states
corresponding to a subgroup with probability $1/d$, e.g. the line state
$\rho_{\alpha_1} = \frac{1}{d} \sum_{k=0}^{d-1} P_{k,0}$ (see also
Fig.\ref{phaseSpace3} and Fig.\ref{phaseSpace4}). This gives rise to a
kernel polytope $\K_d$, which is defined as convex mixture of these separable
line or sublattice states with $d$ elements:
\begin{gather}
\K_d \equiv \lbrace \rho = \sum_\alpha \lambda_{\alpha} \rho_{\alpha}~ |~
\lambda_\alpha \geq 0, \sum_\alpha \lambda_\alpha = 1 \rbrace
\end{gather}
All states in the kernel polytope are convex combinations of separable states,
consequently each state in $\K_d$ is by construction separable and the center
corresponds to the maximally mixed state. \\

\subsection*{Bell diagonal state generation in arbitrary dimension}
\label{sec:bell-diagonal-state}
Using the geometric representation of $\M_d$ in Euclidean space, any state can
be generated by specifying its $d^2$ coordinates $c_{k,l}$. Random sampling as
well as deterministic procedures can be used to generate states. It was shown
\cite{PoppACS} that via random sampling, uniformly distributed states in $\M_d$
can be generated and used to estimate the relative volumes of the entanglement
classes of separable, bound entangled and free entangled states in $\M_3$. The
same method can be used for $d>3$ to generate states in $\M_d$ and $\E_d$ by
drawing the first $d^2-1$ coordinates from a uniform distribution in the range
$[0,1]$ for $\M_d$ or $[0, 1/d]$ for $\E_d$ respectively. The remaining
coordinate is then chosen according to the normalization condition. If this is
not possible with a non-negative probability, the coordinates do not represent a
normalized physical state and is therefore rejected. This form of rejection
sampling becomes less effective with growing dimension $d$ because the
probability of a random state being rejected increases rapidly for $d \geq
5$. However, several methods exist to sample uniformly distributed points on the
standard simplex and thus in $\M_d$ in any dimension (see
Ref.\cite{simplexSampling} and the references therein), but these methods are
not accessible for sampling of states located only in the polytopes $\E_d$ or
$\K_d$.

\subsection*{Symmetries and their generation}
\label{sec:symm-their-gener}
The ring structure of the Weyl operators $W_{k,l}$ can be used to define linear
symmetry transformations \cite{baumgartnerHiesmayr, baumgartner1} acting on
states in $\M_d$. These transformations act as permutations on the Bell basis
projectors $\Pkl$ or equivalently as permutations of the coordinates $c_{kl}$ of
a state in $\M_d$. These symmetry transformations form a group and are known to
conserve both the PPT property and entanglement. Together these properties imply
the conservation of the entanglement class \cite{PoppACS}, meaning that the subsets of
separable, bound entangled and free entangled states are mapped to themselves. For
numerical implementations as well as for understanding their action on a given
state, any linear symmetry $s$ can be characterized by their action on the basis
projectors $s: \Pkl \rightarrow P_{k', l'}$. All elements of this symmetry group
can be generated by combined application of the following group generators (all
mathematical operations on the indices are defined as$\pmod d$):
\begin{itemize}
\item Momentum inversion: $m: \Pkl \rightarrow P_{-k,l}$
\item Quarter rotation: $r: \Pkl \rightarrow P_{k,-l}$
\item Vertical sheer: $v: \Pkl \rightarrow P_{k+l,l}$
\item Translation: $t_{p,q}: \Pkl \rightarrow P_{k+p,l+q}$ for $p, q
  \in (0, \cdots, d-1)$
\end{itemize}
Note that the translation of a Bell diagonal state can also be realized by
applying a corresponding Weyl operator to the mixed state. This is not the case
for the other generators, as their action on each Bell state that contributes to
the mixture depends on that state. Due to the finite number of elements $(k,l)$
in the phase space induced by the Weyl operators $W_{k,l}$, the number of
distinct symmetries generated by the generators above is finite as well and can
be generated numerically. The number of existing symmetries grows quickly with
the dimension. For $d=2$, $24$ distinct symmetries can be generated, for $d=3$,
$432$ and for $d=4$ already $1536$ symmetries of this group exist. Taking these
symmetries into account is essential for the numerical methods to be effective
in the $d^2$ dimensional space.

\subsection*{Criteria for the detection of entanglement and separability}
\label{sec:crit-detect-entangl}
The separability problem to decide whether a given mixed quantum state is
separable or entangled has been shown to be NP-hard with respect to the
dimension $d$ as complexity measure \cite{nphard, nphard-strong} and lacks an
efficient general solution by polynomial in time algorithms, which is also the
case for states in $\M_d$ for general dimension $d$. It is currently unclear,
whether an efficient general solution exists for the separability problem in
$\M_d$. For $d=2$, however, all entangled states can be detected by the PPT
criterion \cite{peres, horodeckiCriterion} and recently, an almost complete
solution was presented \cite{PoppACS} for $d=3$, in the sense that any random
unknown PPT state in $\M_3$ can be classified numerically with a probability of
success of $95\%$. The methods used in this work can be equivalently used or
extended to be applicable for $d>3 $. In the following we introduce those
methods shortly. For more detailed information, the reader is referred to
Ref. \cite{PoppACS} and the references therein. \\
\\
\textbf{E1: PPT criterion} \\
The ``Positive Partial Transpose (PPT)'' or ``Peres-Horodecki'' criterion
\cite{peres} detects entanglement for a bipartite state if it has at least one
negative eigenvalue (in which case it is said to be ``NPT''). For $d=2$ it
detects all entangled states, but for $d \geq 3$ it is only sufficient due to
the existence of PPT- or bound entangled states. The partial transpose $\Gamma$
acts on the basis states of a bipartite state as
$(\ketbra{i}{j} \otimes \ketbra{k}{l})^{\Gamma} \equiv \ketbra{i}{j} \otimes
\ketbra{l}{k}$. \\ \\
\textbf{E2: Realignment criterion} \\
The realignment operation $R$ is defined as
$(\ketbra{i}{j} \otimes \ketbra{k}{l})_R \equiv \ketbra{i}{k} \otimes
\ketbra{j}{l}$. The realignment criterion \cite{realignment} states that if the
sum of singular values of the realigned state $\sigma_R$ are larger than $1$,
then $\sigma$ is entangled. Like the PPT criterion it is only sufficient for
entanglement. Bound entangled states can be detected by this criterion, but it
does not detect all NPT states in general. \\ \\
\textbf{E3: Quasi-pure concurrence criterion} \\
The quasi-pure approximation \cite{BaeQuasipure} $C_{qp}$ of the concurrence
\cite{woottersConcurrence} allows the efficient detection of entanglement
including its bound form. The approximation takes an explicit form for states in
$\M_d$: A state $\rho = \sumkl c_{k,l}P_{k,l} \in \M_d$ is entangled if
$C_{qp}(\rho) = \max(0, S_{nm} - \sum_{(k,l) \neq (n,m)} S_{k,l})>0$ where the
$S_{k,l}$ are explicitly given by \cite{BaeQuasipure}
\begin{gather}
S_{k,l} = \sqrt{
  \frac{d}{2(d-1)} c_{k,l}
  [(1-\frac{2}{d}) c_{n,m} \delta_{k,n} \delta_{l,m}
  + \frac{1}{d^2} c_{(2n-k)mod~d,(2m-l)mod~d}]
}
\end{gather}
and $(n,m)$ is a multi-index of the coordinate of the largest value
$\lbrace c_{k,l} \rbrace $.
\\ \\
\textbf{E4: MUB criterion} \\
A set of orthonormal bases $\lbrace B_k \rbrace$ and
$B_k = \lbrace \ket{i_k} ~|~ i = 0,\ldots, (d-1) \rbrace$ is called ``mutually
unbiased bases (MUB)'' if $\forall k \neq l$:
$ |\braket{i_k}{j_l}|^2 = \frac{1}{d} ~~~ \forall i,j = 0, \ldots, (d-1)$.  At
most $d+1$ MUBs exist \cite{woottersMUB, BandyopadhyayMUB}, in which case it was
shown \cite{bae2021measurements, hiesmayrLoeffler, SpenglerMUB} that the sum of
``mutual predictabilities'' obeys
\begin{gather}
  \label{eq:mub}
  I_{d+1}(\rho_s) = \sum_{k=1}^{d+1} C_k(\rho_s) \leq 2
\end{gather}
for all separable states $(\rho_s)$, when defining
\begin{gather}
  \label{eq:2}
  C_1(\rho) = \sum_{i=0}^{d-1} \bra{i_1}\otimes \bra{(i_1+s)^*} \rho
  \ket{i_1}\otimes \ket{(i_1+s)^*}, \\
  C_k(\rho) =   \sum_{i=0}^{d-1} \bra{i_k}\otimes \bra{i_k^*} \rho \ket{i_k} \otimes
  \ket{i_k^*},~ k=2,\dots,d+1.
\end{gather}
Here, $s = 0,1, \dots, (d-1)$ and $i_k^*$ denotes the complex conjugate vector
element. The MUB criterion thus indicates that if any state violates
(\ref{eq:mub}), it is entangled.  If $s > 0$ and $s \neq d/2$ the MUB criterion
allows the detection of PPT entangled states \cite{bae2021measurements}, which
was also experimentally demonstrated for entangled photons
\cite{hiesmayrLoeffler} in the case $d=3$. Note, that other, inequivalent MUBs
exist, including extendible or unextendible sets of bases that contain less than
$d+1$ elements \cite{ineqMUBs}. The set of entangled states that are detected by
the MUB criterion generally depends on the used MUB. For this work, we use MUBs
of $d+1$ elements as given in the Appendix A1 and set $s=2$ for $d=3$ and $s=3$
for $d=4$.
\\ \\
\textbf{E5: Numerically generated entanglement witnesses}\\
An entanglement witness \cite{EWs} (``EW'') $W$ is an observable which implies
an upper bound $U$ and also a lower bound \cite{EW20} $L$
($U,L \in \mathbb{R}$), for separable states $\rho_s$:
\begin{gather}
  \label{eq:1}
  L \leq \tr[\rho_s W] \leq U
\end{gather}
A state $\rho$ is ``detected by W'' to be entangled, if
$\tr[\rho W] \notin [L,U]$. For the system $ \M_d$, EWs of the form
$W = \sumkl \kappa_{k,l} \Pkl$ with $\kappa_{k,l} \in [-1,1]$ can detect all
entangled states \cite{baumgartner1}. In this case
$\rho = \sumkl c_{k,l}\Pkl \in \M_d$ and
$\tr[\rho W]= \sumkl c_{k,l} \kappa_{k,l} \equiv c \cdot \kappa$ using the
standard scalar product of the $d^2$-dimensional vectors $c$ and $\kappa$ with
coefficients $c_{k,l}$ and $\kappa_{k,l}$. Using the geometric representation of
$\M_d$, an EW defines two $(d^2-1)$-dimensional hyper-planes via
$c_L \cdot \kappa = L$ and $c_U \cdot \kappa = U$ and induced halfspaces. Any
point in the simplex but outside of the intersection of these halfspaces is
entangled. An parameterization of unitaries \cite{spenglerHuberHiesmayr2} can be
used to numerically determine the bounds for any EW defined by its coefficients
$\kappa_{k,l}$ to create EWs for states in $\M_d$ numerically.
\\ \\
Leveraging the geometric characterization of $\M_d$, also sufficient criteria to
detect separable states have been developed and used to analyze $\M_3$
\cite{PoppACS}. They can be applied for $d>3$ as well and are shortly stated
here: \\
\\
\textbf{S1: Extended kernel criterion}\\
The convexity of the set of separable states can be used to check if an unknown
state is contained in the separable hull of known separable states via linear
programming. For this work, the implementation of Ref.\cite{lazysets}, was used
to check an unknown state for separability based on the convex hull of a given
set of separable states. Using known separable states in $\M_d$ as vertices,
they form a polytope which approximates the convex set of all separable states
in $\M_d$. The effectiveness of this criterion depends on the quality of this
approximation. More separable vertices for the polytope to improve the
approximation increase the probability to detect new separable states, but on
the other hand the complexity of the linear program also increases. It is
therefore important to use vertices that are spatially uniformly distributed and
as close to the surface of the set of separable states as possible.  The
sublattice states $\rho_\alpha$ of the kernel polytope $\K_d$ meet those
requirements \cite{baumgartner1} and by using the entanglement-class-conserving
symmetries, more vertices to extend the separable
kernel can be generated.\\
\\
\textbf{S2: Weyl/Spin representation criterion}\\
Based on the Weyl relations \cite{teleportingWeyl}
\begin{gather}
  W_{k_1,l_1}W_{k_2,l_2} = w^{l_1 k_2}W_{k_1+k_2, l_1+l_2}  \\
  W_{k,l}^\dagger = w^{k l}W_{-k, -l} = W_{k,l}^{-1}
\end{gather}
one can see that the Weyl operators form an orthogonal basis in the space of
$d \cross d$ matrices with respect to the trace norm
$ \langle A,B \rangle \equiv \tr[A^\dagger B]$. Representing a density matrix
$\sigma$ as $ \sigma = \frac{1}{d} \sumkl s_{k,l} W_{k,l}$ defines the
coefficients of this ``Weyl representation'' as
$s_{k,l} = \tr[W_{k,l}^\dagger \sigma]$. For bipartite states,
$W_{\mu, \nu} \equiv W_{\mu_1, \nu_1} \otimes W_{\mu_2, \nu_2}$ the
coefficients are indexed as $s_{\mu, \nu}$. It was shown \cite{weylRepCrit} that if
$\sum_{\mu, \nu} |s_{\mu, \nu}| \leq 2$, then $\rho$ is separable. This
criterion for separability is named ``Weyl'' or ``spin representation
criterion''.
\\ \\
\subsection*{Symmetry classification}
\label{symclas}
The group of entanglement class conserving symmetries (see ``Symmetries and
their generation'') provide further methods to determine the entanglement class
of a given unknown state in $\M_d$. First,the set all symmetric states, i.e. the
orbit of the unknown state, is generated by application of the according
transformation for all generated symmetries. Then, this set is analyzed with
respect to the available criteria. If the entanglement class is determined for
one of the symmetric states, then all symmetric states are certainly of the same
class. This method can additionally be used to generate more states of a certain
class for further investigations.

\newpage
\section*{Results}
\label{sec:results}

\subsection*{Volume of PPT states}
\label{sec:volume-ppt-states}
\begin{figure}[H]
  \centering
  \includegraphics[width=\linewidth]{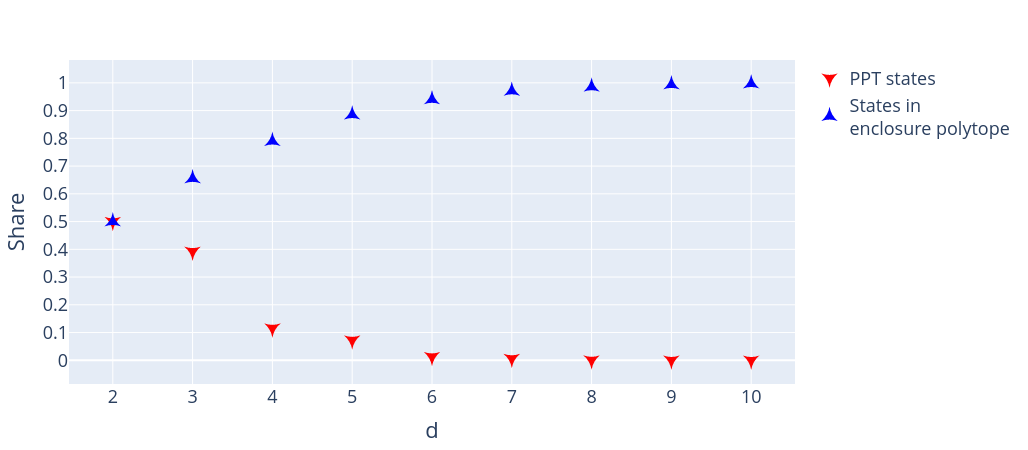}
  \caption{Relative volumes of the enclosure polytope $\E_d$ and PPT states in $\M_d$
    for different dimensions $d$. Figure created with Ref.\cite{plotlyJs}.}
  \label{pptEnc}
\end{figure}

A first application leveraging the presented methods is to determine the
relative volume of states with positive partial transposition in $\M_d$. It was
shown \cite{volOfSep1, volOfSep2} that the volume of general separable and bound
entangled quantum states decreases exponentially with the dimension of the
system. Here, we determine the relative volumes for Bell diagonal states.\\
As described in section "Relevant subsets", all states with positive partial
transposition are necessarily located in the enclosure polytope $\E_d$
(\ref{encP}) when represented in Euclidean space. This property of $\M_d$ yields
an upper bound of the relative share of PPT states in the simplex by comparing
the total volume of $\M_d$ to the volume of $\E_d$. The enclosure polytope
generally contains both PPT and NPT states, but the ratio depends on the
dimension $d$. For $d=2$, all states of $\E_2$ are known to be separable and
thus PPT \cite{geoPic}, no PPT/bound entangled states exist. For $d=3$ it was
numerically shown \cite{PoppACS} that approximately $60.0\%$ of the states in
$\E_3$ ($39\%$ of all states in $\M_3$) are PPT. In order to determine the
relative volumes of PPT, we generate a large number of uniformly distributed
states for dimensions $d=2, \dots, 10$ and check if they are in the enclosure
polytope and if they are PPT. The results are summarized in Figure \ref{pptEnc}.
 \ \\
The analysis demonstrates that despite of the fact that the relative volume of
the enclosure polytope grows with increasing dimension, the relative number of
PPT states quickly decreases. For $d=4$, $11.6\%$ are PPT and the
enclosure polytope $\E_4$ makes up $79.0\%$ of $\M_4$. For $d=5$, the relative
PPT volume reduces to $7.3\%$ and already for $d=6$, less than $1\%$ are PPT,
although $97.1\%$ of states in $\M_6$ are located in $\E_6$.

\subsection*{Entanglement classification of PPT states in $\E_d$}
\label{sec:entangl-class-e_d}

In order to compare the entanglement properties of mixed Bell diagonal states
for the dimension $d=2,3,4$, we use uniformly generated random states in $\E_d$
and classify them with the criteria for separability and entanglement presented
above in order to estimate the share of each
entanglement class. As mentioned in section ``Volume of PPT states'', the
classification of bipartite qubits can be completely characterized by the PPT
criterion (E1): all states in $\E_2$ are PPT and separable, no PPT entangled
states exist. For $d=3$, we take the results of the previous investigation
\cite{PoppACS}, in which $96.1 \%$ of generated states in $\E_3$ have been
successfully classified. To determine the relative volumes of entanglement
classes among the PPT states for $d=4$ with comparable precision, $40000$ random
states are generated
in $\E_4$ out of which $96.7\%$ can be successfully classified. \\
The $60 \%$ of states in $\E_3$ and $14.6 \%$ of $\E_4$ that have positive
partial transposition are labeled as ``SEP'' or ``BOUND'' if they are detected
as separable or entangled by the criteria E2-E5 or S1-S2. If none of the
criteria allows classification, the state is labeled ``PPT-UNKNOWN''. Table
\ref{classShares} summarizes the results:
\begin{table}[H]
  \centering
  \begin{tabular}{|l|r|r|r|}
    \hline
    \textbf{  Entanglement Class  } &
                                      \textbf{  Share of PPT for $d=2$ } &
                                                                           \textbf{  Share of PPT for $d=3$ } &
      \textbf{  Share of PPT for $d=4$ } 
    \\
    \hline
    SEP & $100 \%$ & $81.0 \%$ & $75.7 \%$ \\
    \hline
    BOUND & $0 \%$ & $13.9 \%$ & $1.7 \%$ \\    
    \hline
    PPT-UNKNOWN & $0 \% $ & $5.1 \%$ & $22.6 \%$\\
    \hline
  \end{tabular}
  \caption{Relative volumes of entanglement classes among the PPT states for $d=2,3,4$}
  \label{classShares}
\end{table} \ \\
Comparing the numerical classifications for $d=3$ and $d=4$, three noteworthy
differences can be seen: First, the relative number of PPT-UNKNOWN states is
significantly higher for $d=4$ ($22.6\%$) than for $d=3$ ($5.1\%$), although the
success rates ($96.7 \%$ and $96.1 \%$) for the classification with respect to
the total number of generated states in $\E_4$ and $\E_3$ are similar. Second,
the share of separable states in $d=4$ ($75.7\%$) among the PPT states is quite
high, in spite of the large number of yet to be classified PPT states. Third,
the number of detected bound entangled states in $d=4$ ($1.7 \%$) is
significantly lower than for $d=3$ ($13.9 \%$). Although it is possible that a
large part of the PPT-UNKNOWN states are in fact BOUND and thus could
potentially be detected by criterion E5, this shows that the detection
capability of the analytical criteria (E2-E4) is more limited for $d=4$ than
for $d=3$.

\subsection*{Detection capabilities and relations of applied criteria}
\label{sec:detect-capab-relat}
The detection and thus differentiation between bound entangled and separable states
is the core of the separability problem. Hence, the detection capabilities of
the presented detectors for the classes SEP and BOUND are of special
interest. Table \ref{tab:bound-sep-detectors} shows for each relevant criterion
and dimension $d$ the share of detected states among all SEP, respectively
BOUND, classified states.

\begin{figure}[H]
  \centering
  \includegraphics[width=1\linewidth]{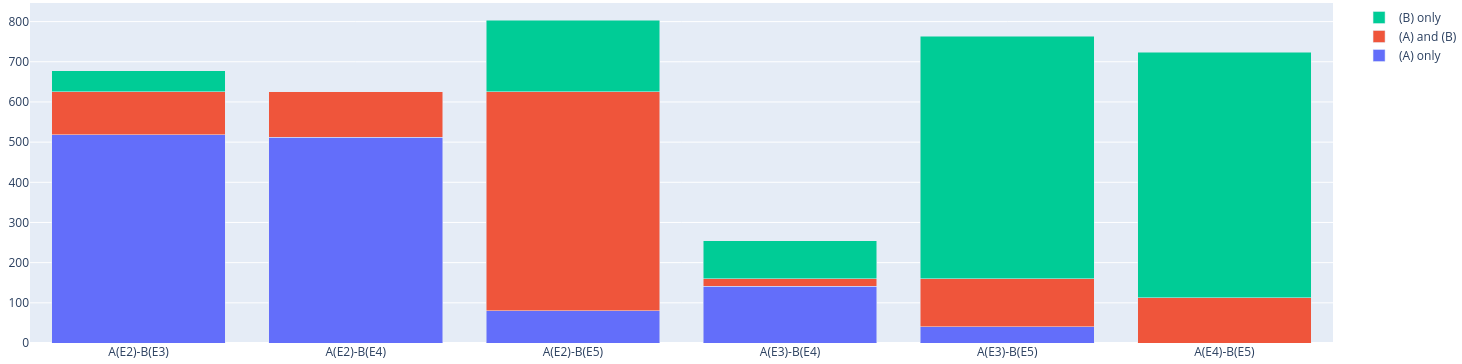}
  \caption{Pairwise comparison of number of exclusively (blue and green) and
    jointly (red) detected states for $d=3$. Figure created with
    Ref.\cite{plotlyJs}.} 
  \label{inter3}
\end{figure}

\begin{figure}[H]
  \centering
  \includegraphics[width=1\linewidth]{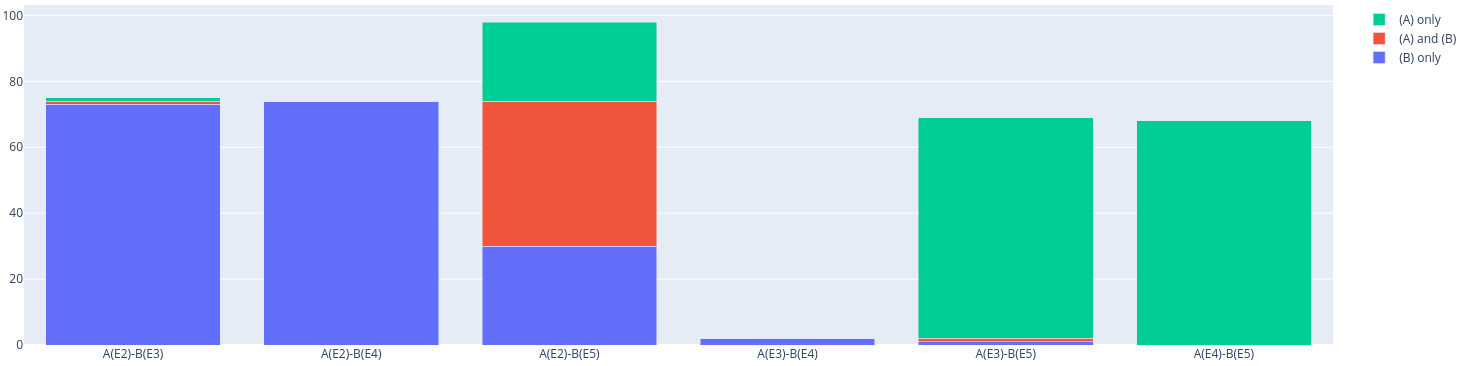}
  \caption{Pairwise comparison of number of exclusively (blue and green) and
    jointly (red) detected states $d=4$. Figure created with
    Ref.\cite{plotlyJs}.} 
  \label{inter4}
\end{figure}

\begin{table}[H]
\centering
\begin{tabular}{|l|l|r|r|}
  \hline
  \textbf{Entanglement class} &
  \textbf{Criterion} &
  \textbf{Share in class for $d=3$} &
  \textbf{Share in class for $d=4$}  \\
  \hline
  SEP & S1  & $100 \%$ & $100 \%$ \\
  \hline
  SEP & S2 & $1.7 \%$ & $0 \%$ \\
  \hline
  BOUND & E2 & $74.9 \%$ & $74.7 \%$ \\
  \hline
  BOUND  & E3 & $19.1 \%$ & $2.2 \%$ \\
  \hline
  BOUND & E4 & $13.5 \%$ & $0 \%$ \\
  \hline
  BOUND & E5 & $86.6 \%$ & $68.7 \%$\\
  \hline
\end{tabular}
\caption{BOUND and SEP detectors and their detection shares for $d=3$ and $d=4$.}
\label{tab:bound-sep-detectors}
\end{table} \ \\
On the one hand, one notices that the strongest detectors for bound entangled
Bell diagonal qutrits are also the most successful detectors in $d=4$, namely E2
and E5. Relative to the total amount of detected bound entangled states, E2
seems to perform equally well in both dimensions. However, due to the large
amount of PPT-UNKNOWN states, the relative detection power could be much
worse. Still, E2 is clearly the strongest applied analytical criterion for $d=3$
and $d=4$. The criterion based on combining many numerically generated
witnesses, E5, detects a large share of the identified BOUND states for both
analyzed dimensions, however, the share is lower for $d=4$ ($86.6 \%$ for $d=3$,
$68.7 \%$ for $d=4$). Considering the PPT-UNKNOWN states, the true detection
capability of this criterion might be even below the determined share of
$68.7 \%$ for $d=4$, even though more EWs were used for $d=4$ (approximately
$22700$ compared to $16700$ for $d=3$). This indicates that a single randomly
generated numerical EW most likely is a weaker detector for the higher dimension. \\
On the other hand, the other detectors are clearly weaker in $d=4$ compared to
$d=3$. The second strongest criterion in $d=3$, E3, detects $19.1 \%$ of the
BOUND qutrits in, while only $2.2 \%$ in $d=4$. Again, the true detection
capability is likely even below that due to the large number of unclassified PPT
states. E4 detects a significant share ($13.5 \%$) of bound entangled qutrits
while no PPT entangled qudits for $d=4$. Likewise, S2 detects no states as
separable for $d=4$.\\
These differences are also clearly reflected when comparing the criteria E2-E5
pairwise as shown in Figures \ref{inter3} and \ref{inter4}. The only criteria
that have a significant number of jointly detected states in $d=4$ are the
detectors E2 and E5 ($44.4 \%$ of combined detected states), although the share
is smaller than for $d=3$ ($67.7 \%$), confirming the reduced effectiveness of
numerical EWs in the higher dimension. For $d=4$, the other pairs are rather
trivial, because of the very low number of detected states by E3 and E4. It
should be noted, however that E3 detects one bound state that is neither
detected by E2 nor by E5. Interestingly, this criterion also detects significant
shares of bound entanglement that are not detected jointly by E2 or E5 in
$d=3$. \\
A final remark can be made related to the purity $\tr \rho^2$ of detected bound
entangled states $\rho$. For $d=3$, the least pure states were detected by the
criterion E3 but not E2 or E4. The few detected states for $d=4$ do not allow to
confirm this observation, although it can be noted that the least pure bound
state is also uniquely detected by E3.

\subsection*{Visual analysis of separable and bound entangled states in $\M_3$ and $\M_4$}
\label{sec:visu-analys-separ}
\begin{figure}[H]
  \centering
  \begin{subfigure}{.4\textwidth}
    \includegraphics[width=0.8\linewidth]{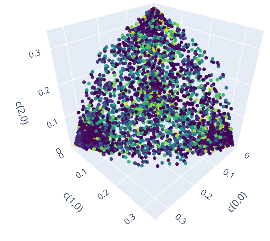}
  \end{subfigure}%
  \begin{subfigure}{.6\textwidth}
    \includegraphics[width=\linewidth]{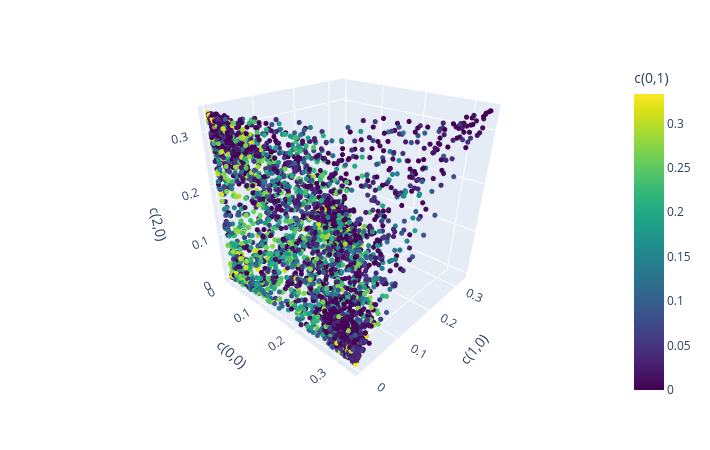}
  \end{subfigure}
  \caption{$(c_{0,0}, c_{1,0}, c_{2,0}, c_{0,1})$ for optimized separable states
    for $d=3$, demonstrating strong correlations of on-line coordinates. Figure
    created with Ref.\cite{plotlyJs}.} 
  \label{sepOptLine3}
\end{figure}

\begin{figure}[H]
  \centering
  \begin{subfigure}{.5\textwidth}
    \centering
    \includegraphics[width=0.7\linewidth]{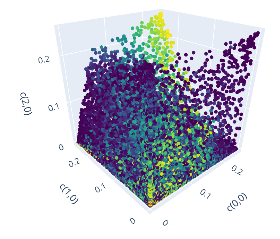}
  \end{subfigure}%
  \begin{subfigure}{.5\textwidth}
    \centering
    \includegraphics[width=\linewidth]{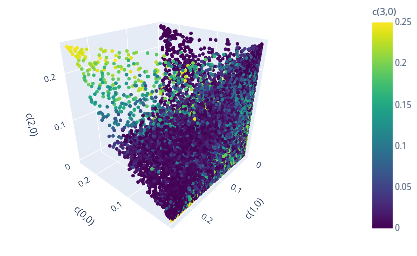}
  \end{subfigure}
  \caption{$(c_{0,0}, c_{1,0}, c_{2,0}, c_{3,0})$ for optimized separable states
    for $d=4$, demonstrating strong correlations of on-line and
    additionally on-sublattice coordinates. Figure created with
    Ref.\cite{plotlyJs}.} 
  \label{sepOptLine4}
\end{figure}

\begin{figure}[H]
  \centering
  \begin{subfigure}{.5\textwidth}
    \centering
    \includegraphics[width=.7\linewidth]{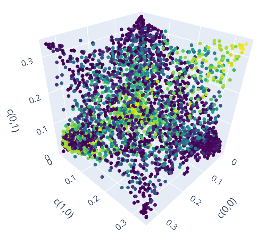}
  \end{subfigure}%
  \begin{subfigure}{.5\textwidth}
    \centering
    \includegraphics[width=\linewidth]{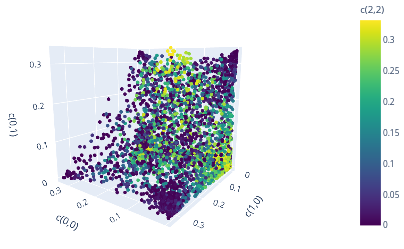}
  \end{subfigure}
  \caption{$(c_{0,0}, c_{1,0}, c_{0,1}, c_{2,2})$ for optimized separable states
    for $d=3$, demonstrating necessity of full-line mixing for separability of
    states that concentrate most probability on $3$ Bell states. Figure created
    with Ref.\cite{plotlyJs}.}
  \label{sepOpt2onlLine3}
\end{figure}

\begin{figure}[H]
  \centering
  \begin{subfigure}{.5\textwidth}
    \centering
    \includegraphics[width=.7\linewidth]{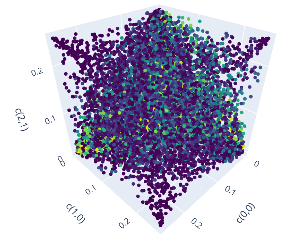}
  \end{subfigure}%
  \begin{subfigure}{.5\textwidth}
    \centering
    \includegraphics[width=\linewidth]{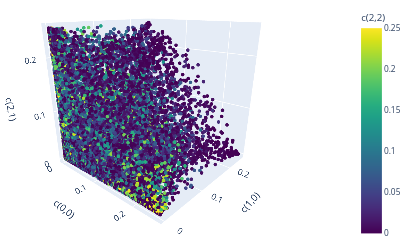}
  \end{subfigure}
  \caption{$(c_{0,0}, c_{1,0}, c_{2,1}, c_{2,2})$ for optimized separable states
    for $d=4$, demonstrating necessity of full-line mixing for separability of
    states that concentrate most probability on $4$ Bell states. Figure created
    with Ref.\cite{plotlyJs}.}
  \label{sepOpt3onlLine4}
\end{figure}

\begin{figure}[H]
  \centering
  \begin{subfigure}{.5\textwidth}
    \centering
    \includegraphics[width=.7\linewidth]{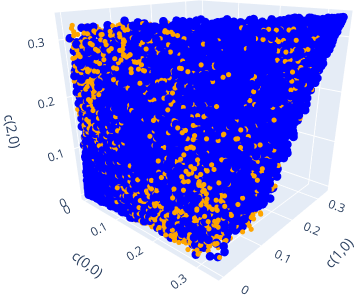}
  \end{subfigure}%
  \begin{subfigure}{.5\textwidth}
    \centering
    \includegraphics[width=.7\linewidth]{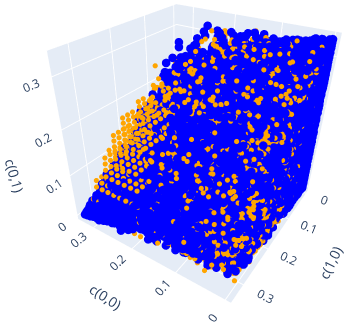}
  \end{subfigure}
  \caption{Comparison of projections to on-line (left) vs. off-line (right)
    coordinates for separable (blue) and bound entangled (orange) states in
    $d=3$. Off-line projections show weaker geometric restrictions for bound
    entangled states in certain regions. Figure created with Ref.\cite{plotlyJs}.  }
  \label{sepVsBound}
\end{figure}

In this section we analyze and compare the structure of the set of separable and
PPT entangled Bell diagonal states in dimension $d=3$ and $d=4$. Analyzing the
geometric properties of those sets helps to improve geometric methods of
entanglement detection or to provide new insights about the geometry of such
quantum states in general (see e.g. Ref.\cite{geoQS}). For $d=2$, the geometric
properties have been analyzed in detail before \cite{geoPic}. Here, we visualize
$d$ dimensional projections of PPT states in $\M_d$ to demonstrate clear patterns
that relate the geometric structure of the set of separable or bound entangled
states to the algebraic structure of the Weyl operators and related Bells
states. In particular, we make two qualitative observations:

\begin{itemize}
\item (Fig.\ref{sepOptLine3} - Fig.\ref{sepOpt3onlLine4}):\\
    The set of separable states in $\M_d$ is geometrically strongly restricted
    by the $d$ element subgroup structure of the Weyl operators. More precisely,
    the more probability of a separable mixed state is concentrated on $d$ Bell
    states, the closer it needs to be to a subgroup state (for definition see
    Fig.\ref{phaseSpace3}, Fig.\ref{phaseSpace4} and related discussion).
\item (Fig.\ref{sepVsBound}): \\
    Similar restrictions related to this subgroup structure are
    present for bound entangled states, but certain restrictions are stronger
    for separable states than for bound entangled states. This difference can
    potentially be leveraged to detect bound entanglement.
\end{itemize}
The set of separable states in $\M_d$ forms a convex geometric body when
represented as set of points in $d^2$ dimensional Euclidean space via their
coordinates $\ckl$. Naturally, not all coordinates of the $d^2$ dimensional
space can be visualized at once, however, the symmetries (see section
``Symmetries and their generation'') allow to capture some essential geometric
properties, even if only $d$ coordinates are shown. The reason for this are
$d$-element subgroups that are induced by the underlying ring structure of the
Weyl operators (see Fig.\ref{phaseSpace3} and \ref{phaseSpace4}). Subgroups can
always be mapped onto each other with a corresponding symmetry transformation
\cite{baumgartner1}. Since these transformations conserve the entanglement class
and act as permutations on the coordinates $\ckl$, different sets of $d$
coordinates will show the same geometric properties if they are symmetric. Here,
we consider the special subgroups represented by lines \cite{baumgartner1,
  baumgartnerHiesmayr} in the discrete phase space that are induced by the
consecutive application of simple translations $t_{p,q}$. For example in $d=3$,
the indices $(k,l)$ of the first three coordinates $(c_{0,0}, c_{1,0},c_{2,0})$
form a line (red markers in Fig.\ref{phaseSpace3}), because they are related by
the translation $t_{1,0}$. The geometric properties of these coordinates are
then equivalent to e.g. those of $(c_{0,0},c_{1,1},c_{2,2})$ (blue markers in
Fig.\ref{phaseSpace3}), as a suitable symmetry transformation relates the
collections.
\\
For visualizations we use separable states, which are specifically optimized to
be close to the surface of the set of separable states, and random samples of
PPT entangled states. We use three coordinates for a 3D-visualization and encode
a fourth coordinate by color. Note that all PPT states are contained in the
enclosure polytope, so it suffices to limit the range of the coordinates
to $[0,1/d]$. \\
To demonstrate the first observation, we compare the projections of separable
states for $d=3$ and $d=4$ to $d$ coordinates that relate either all to the same
$d$-element subgroup or not. Fig.\ref{sepOptLine3} shows the geometric
distribution of the first three coordinates on a line and a fourth coordinate
encoded by the color for $d=3$ from two point of views. A structure similar to
a cone spanned by the corners of the enclosing polytope $\E_3$
$\lbrace (1/3,0,0), (0,1/3,0),(0,0,1/3) \rbrace$ and the subgroup state
$\lbrace (1/3, 1/3, 1/3) \rbrace$ is visible, while no correlation with the
off-line coordinate $c_{0,1}$ can be identified. Note that there are no
separable states, for which two coordinates of the line, i.e.. $c_{0,0}$ and
$c_{1,0}$, are large, while the remaining line coordinate, i.e..  $c_{2,0}$, is
small.

Similar observations can be made for $d=4$, when projecting to the on-line
coordinates $(c_{0,0}, c_{1,0}, c_{2,0}, c_{3,0})$ in
Fig.\ref{sepOptLine4}. Several correlations between the coordinates can be
seen. Corresponding to the yellow cone, we see an accumulation of separable
states that have high mixing probabilities for all of the four Bell states on
the line. The blue cone pointing to the point $(1/4,0, 1/4)$, on the other
hand, relates to a different subgroup with indices
$\lbrace (0,0), (2,0), (0,2), (2,2) \rbrace$ (red markers in
Fig.\ref{phaseSpace4}).  Finally, there is no symmetric cone and thus no
separable states in the vicinity of $(0, 1/4, 1/4)$. Note that there is also no
corresponding subgroup that contains the indices $(1,0)$ and $(2,0)$ but
excludes the remaining line elements $(0,0)$ and $(3,0)$.
The visualizations above demonstrate that for each $d$-element subgroup, there
is an accumulation of separable states in the vicinity of the
corresponding subgroup state. \\
Consider now that $d$ Bell states share most of the probability of a mixed
state. We demonstrate that if those Bell states do not all correspond to the
same subgroup, then the mixed state cannot be separable. First, note that if all
probability is concentrated on less than $d$ states, than at least one
probability must exceed $1/d$, in which case the state cannot be PPT and
therefore must be (NPT) entangled (see ``Enclosure polytope'' on p.4). Below, we
visualize projections to coordinates, of which two relate to the same line and
remaining coordinates are chosen to be part of different lines. Consider for
$d=3$ the projection to the line coordinates defined by $(c_{0,0}, c_{1,0})$
together with the off-line coordinates $c_{0,1}$ and $c_{0,2}$
(Fig.\ref{sepOpt2onlLine3}). No separable states are present for large values of
both on-line coordinates $(c_{0,0}, c_{1,0})$ and large off-line values for
$c_{0,1}$ or $c_{2,2}$.
The same observation can be made for $d=4$. Choosing again two coordinates
$(c_{0,0}, c_{1,0})$ to define a subgroup and two coordinates that are not part of
it, one sees a similar structure in Fig.\ref{sepOpt3onlLine4}. Again, if the
values of the line coordinates $c_{0,0}$ and $ c_{1,0}$ are large, there is
no separable state for large values of the off-line coordinates $c_{2,1}$ or
$ c_{2,2}$.
 \ \\
Due to the discussed symmetries, these characterizations hold for all
coordinates that have a similar relation regarding their corresponding
subgroups. We have therefore demonstrated the first observation: Considering
mixed states that concentrate most of the probability on $d$ Bell states,
similarity to the subgroup states is necessary for separability. Most mixed
separable states have their probabilities distributed on all Bell states and are
therefore centered around the maximally mixed state. In this case they are
likely contained in the convex hull of the subgroup states that define the
kernel polytope. We have therefore shown that this linear approximation also
captures the relevant geometrical structure of the whole set of separable states
in $\M_d$. In addition, the observed restrictions allow to construct more
effective approximations, which focus on the areas, in which the surface of the
set of separable states is curved (see e.g. Fig.\ref{sepOptLine3} (left)) and
the linear approximation fails. This can help to improve existing methods of
detecting separability (e.g. the extended kernel criterion S1).
\\ 
We conclude this section by demonstrating the second observation, which states
that the geometric restrictions induced by the subgroup structure of the Weyl
operators are also present for bound entangled states, but can be
distinguished from those for separable states in certain cases.\\
Consider $d=3$, for which a significant amount of bound entangled states can be
classified. In Fig.\ref{sepVsBound}, we again visualize $d$ dimensional
projections of separable states but also include the detected bound entangled
states. Here, we show separable states in blue and bound entangled states in
orange. The graphic on the left of Fig.\ref{sepVsBound} shows the projection to
the coordinates $(c_{0,0}, c_{1,0}, c_{2,0})$ that relate to the same subgroup
(line). One can see that the projections of the bound entangled states are
restricted to the same area as the separable states (also compare
Fig.\ref{sepOptLine3} (right)). The described dependence of on-line projections
on the related subgroups seems therefore to
be a feature of all PPT states.\\
On the right hand, the visualization of the projection to three coordinates
$(c_{0,0}, c_{1,0}, c_{0,1})$, that do not all belong to the same line, shows a
relevant difference. The general form of visualized PPT states is still
dominated by the convex combination of subgroup states (compare
Fig.\ref{sepOpt2onlLine3}). There are no bound entangled states that concentrate
most of the mixing probabilities on three Bell states that do not belong to the
same subgroup. Crucially, however, there is a clearly visible region, in which
only bound entangled states are present. For given on-line coordinate values
$(c_{0,0}, c_{1,0})$, this region is characterized by higher off-line
coordinate values $c_{0,1}$ than those accessible for separable projections.
 \ \\
For $d=3$, this demonstrates the second observation: On the one hand, the
subgroup structure imposes restrictions on all PPT states that determine the
dominant shape of $d$ dimensional projections of separable and bound entangled
states. On the other hand, there exist $d$ dimensional projections under which a
subset of bound entangled states is mapped to regions, which do not contain any
projected separable states. In principle, knowledge about these regions can be
used to construct suitable entanglement witnesses of rank $d$ to detect those
PPT entangled states in $d^2$ dimensions.\\
For $d=4$ the number of classified bound entangled states is not large enough to
confirm similar characteristics, although none of the classified bound entangled
states indicate a qualitative difference to the observations made for $d=3$.

\newpage
\section*{Discussion and Conclusion}
\label{sec:discussion}
In this work we analyzed states of the bipartite system of mixed Bell states,
which are related by Weyl transformations, with focus on subsystems with
dimension $2,3$ and $4$. The entanglement class of these locally maximally mixed
states depends on the mixing probabilities and can be separable, NPT/free
entangled or for $d > 2$ also PPT/bound entangled.\\
In order to investigate the properties for $d=4$, i.e. bipartite ququarts, we
extended the applied methods recently used to analyze the system of bipartite
qutrits \cite{PoppACS}. Leveraging a geometric representation, a random sampling
of uniformly distributed states can be used to estimate the relative sizes of
the entanglement classes via various criteria to detect separability and
entanglement, including its bound version. Using this representation together
with a group of entanglement class preserving symmetries, related analytical
properties and an efficient parameterization\cite{spenglerHuberHiesmayr2} of
states allows us to draw several conclusions about the entanglement properties
of bipartite qudits and their dependence on the dimension for $d=2,3$ and
$4$. \\
A first observation is that the relative number of states with positive partial
transposition, so either separable of bound entangled states, decreases very
quickly with growing dimension, despite of increasing relative volume of the
enclosure polytope $\E_d$, known to contain all PPT states. The share of PPT
states in the full ``magic simplex'' $\M_d$ decreases from $50\%$ for $d=2$ to
$39\%$ for d=3 to $12\%$ for $d=4$. For $d>5$, less than
$1\%$ of states are PPT. \\
The second observation is that significantly less bound entangled states can be
detected, while the number of states that cannot be classified is considerably
higher for $d=4$ than for $d=3$. To determine the relative volumes of
entanglement classes for the enclosure polytope in $d=4$ and compare them to
$d=3$, $40000$ states in $\E_4$ have been created and classified with a
probability of success of $96.7 \%$. In principle, the probability of success
could further be improved by extension of the numerical analyses and more states
could easily be classified. In order to compare to $d=3$, however, the extend of
the numerical analysis and the number of states are chosen to result in a
similar probability of success and number of PPT states. Limited to the set of
PPT states, $77.4 \%$ of states could be successfully differentiated between
separable and bound entangled states. The developed methods can be efficiently
and repeatedly applied to new unknown states to solve the NP-hard ``separability
problem'' with a probability of success of $77.4 \%$ for Bell diagonal ququarts
in the magic simplex. Out of all PPT states in the system $\M_4$, at least
$75.7\%$ are determined to be separable and $1.7\%$ are classified as bound
entangled. The share of detected bound entangled states is clearly smaller than
for $d=3$ ($13.9 \%$), however, compared to the results of $d=3$, a higher share
($22.6$ \% vs $5.1\%$) of PPT states could not be classified and it remains
unclear whether they are separable or bound entangled. \\
A third result can be stated regarding the detection capabilities of the applied
criteria. The applied detectors for separability (S1, S2) or bound entanglement
(E2-E5) are either based on deterministic, analytical conditions (S2, E2, E3,
E4) or on a combined collection of numerically generated objects, i.e., vertices
for the extension of the kernel polytope for S1 or EW-defining hyperplanes for
E5. It can be seen from the large share of unclassified PPT states, as well as
from the relative shares of detected states by the criteria in each class, that
both types of detectors are less powerful for classification in $d=4$. Out of
the analytical criteria for entanglement detection (E2-E4), for $d=4$, only E2
detects a significant amount of bound entangled states, while E3 detects very
few and E4 none at all, though it is known that E4 can detect bound entangled
Bell diagonal states in $d=4$ \cite{bae2021measurements, ineqMUBs}. This is in
strong contrast to $d=3$, where the later two criteria allow the detection of
$19.1\%$ and $13.5 \%$ of all BOUND classified states. Interestingly, E3 still
detects bound entanglement in $d=4$ that cannot be detected with E2. This can
also be observed for $d=3$, where E3 can detect more strongly mixed entangled
states than E2. The numerical criteria S1 and E5 also show reduced detection
capability. Although S1 detects more than $75\%$ of the PPT states as separable,
the large number of unclassified states suggests that many separable states
might not be detected by the used kernel extension. In addition to the share of
classified states, the reduced number of states that are both detected by the
analytical criterion E2 and the numerical E5 is also an indication of lower
detection power of a single randomly generated EW for $d=4$. another striking
difference between $d=3$ and $d=4$ is that S2 does not detect or is very
unlikely to detect any separability for $d=4$.\\
Many BOUND states that are detected by E2 are thus not confirmed by the
criterion E5, which clearly shows that the number of generated EWs is not high
enough, although more EWs were used than for $d=3$. Two main reasons are likely
responsible for the weaker performance in $d=4$: First, the higher dimension of
the Euclidean space and second, the different geometric properties of the set of
separable states in $\M_d$ related to the properties of the Weyl operators and
their induced phase space in non-prime dimensions. Both criteria represent
approximations of this convex set: S1 represents an polytope approximation from
within by identifying separable vertices close to the surface of separable
states, while E5 represents an enclosing approximation with the hyperplanes
defined by the upper and lower bounds of the EWs. The higher the dimension of
the Euclidean space, the more objects (vertices/hyperplanes) are needed for a
sufficient approximation and a generated set of objects may not be sufficient to
achieve a comparable probability of success. The geometric properties of the
(unknown) convex body formed by separable states are also relevant, as they
determine the results of optimization procedures over the set of separable
states in the whole Hilbert space, on which the generation of EWs and
separable vertices rely.\\
Finally, we used classified separable and bound entangled states to enable
visual analyses concerning the structure of PPT states in $\M_d$. We argued that
relevant information can be extracted by considering projections to $d$
coordinates, due to the special symmetries in $\M_d$ for $d=3$ and $d=4$. Two
main qualitative observations were made that relate the structure of separable
and bound entangled states to the algebraic subgroup structure of the Weyl
operators.
\\
On the one hand, it was shown that the states defined by the induced subgroup
structure of the Weyl operators, which are used to define the kernel polytope
$\K_d$, also determine the dominant geometric shape of the body of separable
states. It was demonstrated for both $d=3$ and $d=4$ that a mixed state that
concentrates most of the probability on $d$ Bell states is separable, only if
all Bell states relate to the same subgroup. This insights can be used to
construct better approximations of the set of separable states and thus to
improve methods to detect separability among PPT states.
\\
On the other hand, we have demonstrated that similar restrictions related to the
subgroups hold as well for the $d$ dimensional projections of bound entangled
states. Importantly, however, for some projections, these restrictions seem to
be less strict for bound entangled states than those imposed on separable
states. As a consequence, there are $d$ dimensional areas, which are not
reachable for projections of separable states. In principle, PPT states that are
projected to these areas can be classified as entangled by geometric criteria
that are defined in $d$ dimensional space, instead of the full $(d^2-1)$
dimensional Hilbert space.
\\
In conclusion, the methods for state generation and entanglement analysis
applied to the system of $d=3$ can also be successfully applied to $d=4$ (with
extensions), although with reduced effectiveness. Nonetheless, the presented
methods solve the separability problem for $\M_4$ to a large extend, since any
unknown state in $\M_4$ can efficiently be classified as separable or (bound)
entangled with high probability of success. Significant differences in the
relative volumes of the entanglement classes and in the detection capabilities
of criteria for separability and entanglement are observed for $\M_3$ and
$\M_4$. Relating the algebraic structure of the Weyl operators to the geometry
of $\M_d$, qualitative observations could be made that characterize the
structure of PPT states and propose new potential criteria to detect bound
entanglement in $\M_d$.  These contributions can serve as starting point to
further improve the methods for classification and the general understanding of
the entanglement structure of Bell diagonal qudits or general quantum states. On
the one hand, further numerical investigations, extending the current results in
terms of higher dimensions $d \geq 5$ or the structure of bound entangled
quantum states, are possible. The applied sampling methods remain efficient for
Bell diagonal states and thus allow the confirmation of results concerning the
exponential decrease of volume for separable general quantum states with growing
dimension as reported in Ref.\cite{volOfSep1, volOfSep2} or the observation of
special properties for Bell diagonal states. Recently, a sequentially
constrained Monte Carlo sampler (SCMCS) for quantum states was proposed
\cite{SCMCS}, which allows efficient sampling of quantum states subject to
constraints like PPT or detection properties for specific entanglement
criteria. This method could be used to generate general bound entangled quantum
states and compare their properties to those of Bell diagonal states or for the
detailed investigation of detection capabilities of certain entanglement
witnesses. On the other hand, the reported structures of separable and bound
entangled Bell diagonal states in $\M_d$ indicate properties that can be related
to those of the Weyl operators and their induced phase space structure. The
presented methods help to create, confirm or refute hypothesis about the
structure of separable or entangled Bell diagonal states for different
dimensions. Thus, they could provide a new accesses to the separability problem,
the detection of bound entanglement or application relevant properties of
quantum systems using entangled qudits.

\newpage

\section*{Data availability statement}
All analyzed datasets were generated during the current study and are available
from the corresponding author on reasonable request. \\
The software used to generate the reported results is published as registered
open source package ``BellDiagonalQudits.jl'' \cite{BellDiagonalQudits} available at
\url{https://github.com/kungfugo/BellDiagonalQudits.jl}.

\bibliography{references}

\begin{thebibliography}{10}
\expandafter\ifx\csname url\endcsname\relax
  \def\url#1{\texttt{#1}}\fi
\expandafter\ifx\csname urlprefix\endcsname\relax\def\urlprefix{URL }\fi
\providecommand{\bibinfo}[2]{#2}
\providecommand{\eprint}[2][]{\url{#2}}

\bibitem{distQuantComp}
\bibinfo{author}{Cirac, J.~I.}, \bibinfo{author}{Ekert, A.~K.},
  \bibinfo{author}{Huelga, S.~F.} \& \bibinfo{author}{Macchiavello, C.}
\newblock \bibinfo{title}{Distributed quantum computation over noisy channels}.
\newblock \emph{\bibinfo{journal}{Phys. Rev. A}} \textbf{\bibinfo{volume}{59}},
  \bibinfo{pages}{4249--4254} (\bibinfo{year}{1999}).
\newblock \urlprefix\url{https://link.aps.org/doi/10.1103/PhysRevA.59.4249}.

\bibitem{QuantCryptBell}
\bibinfo{author}{Ekert, A.~K.}
\newblock \bibinfo{title}{Quantum cryptography based on bell's theorem}.
\newblock \emph{\bibinfo{journal}{Phys. Rev. Lett.}}
  \textbf{\bibinfo{volume}{67}}, \bibinfo{pages}{661--663}
  (\bibinfo{year}{1991}).
\newblock \urlprefix\url{https://link.aps.org/doi/10.1103/PhysRevLett.67.661}.

\bibitem{quantSimReview}
\bibinfo{author}{Georgescu, I.~M.}, \bibinfo{author}{Ashhab, S.} \&
  \bibinfo{author}{Nori, F.}
\newblock \bibinfo{title}{Quantum simulation}.
\newblock \emph{\bibinfo{journal}{Rev. Mod. Phys.}}
  \textbf{\bibinfo{volume}{86}}, \bibinfo{pages}{153--185}
  (\bibinfo{year}{2014}).
\newblock \urlprefix\url{https://link.aps.org/doi/10.1103/RevModPhys.86.153}.

\bibitem{twoStepQuantDirCom}
\bibinfo{author}{Deng, F.-G.}, \bibinfo{author}{Long, G.~L.} \&
  \bibinfo{author}{Liu, X.-S.}
\newblock \bibinfo{title}{Two-step quantum direct communication protocol using
  the einstein-podolsky-rosen pair block}.
\newblock \emph{\bibinfo{journal}{Phys. Rev. A}} \textbf{\bibinfo{volume}{68}},
  \bibinfo{pages}{042317} (\bibinfo{year}{2003}).
\newblock \urlprefix\url{https://link.aps.org/doi/10.1103/PhysRevA.68.042317}.

\bibitem{oneStepQSDC}
\bibinfo{author}{Sheng, Y.-B.}, \bibinfo{author}{Zhou, L.} \&
  \bibinfo{author}{Long, G.-L.}
\newblock \bibinfo{title}{One-step quantum secure direct communication}.
\newblock \emph{\bibinfo{journal}{Science Bulletin}}
  \textbf{\bibinfo{volume}{67}}, \bibinfo{pages}{367--374}
  (\bibinfo{year}{2022}).
\newblock \urlprefix\url{https://doi.org/10.1016/j.scib.2021.11.002}.

\bibitem{nielsen}
\bibinfo{author}{Nielsen, M.~A.} \& \bibinfo{author}{Chuang, I.~L.}
\newblock \emph{\bibinfo{title}{Quantum Computation and Quantum Information}}
  (\bibinfo{publisher}{Cambridge University Press}, \bibinfo{year}{2000}).

\bibitem{epr}
\bibinfo{author}{Einstein, A.}, \bibinfo{author}{Podolsky, B.} \&
  \bibinfo{author}{Rosen, N.}
\newblock \bibinfo{title}{Can quantum-mechanical description of physical
  reality be considered complete?}
\newblock \emph{\bibinfo{journal}{Phys. Rev.}} \textbf{\bibinfo{volume}{47}},
  \bibinfo{pages}{777--780} (\bibinfo{year}{1935}).
\newblock \urlprefix\url{https://link.aps.org/doi/10.1103/PhysRev.47.777}.

\bibitem{bellEpr}
\bibinfo{author}{Bell, J.~S.}
\newblock \bibinfo{title}{On the einstein podolsky rosen paradox}.
\newblock \emph{\bibinfo{journal}{Physics Physique Fizika}}
  \textbf{\bibinfo{volume}{1}}, \bibinfo{pages}{195--200}
  (\bibinfo{year}{1964}).
\newblock
  \urlprefix\url{https://link.aps.org/doi/10.1103/PhysicsPhysiqueFizika.1.195}.

\bibitem{loopholeFreeBell}
\bibinfo{author}{Hensen, B.} \emph{et~al.}
\newblock \bibinfo{title}{Loophole-free bell inequality violation using
  electron spins separated by 1.3 kilometres}.
\newblock \emph{\bibinfo{journal}{Nature}} \textbf{\bibinfo{volume}{526}}
  (\bibinfo{year}{2015}).

\bibitem{Moskal_2016}
\bibinfo{author}{Moskal, P.} \emph{et~al.}
\newblock \bibinfo{title}{Time resolution of the plastic scintillator strips
  with matrix photomultiplier readout for j-{PET} tomograph}.
\newblock \emph{\bibinfo{journal}{Physics in Medicine and Biology}}
  \textbf{\bibinfo{volume}{61}}, \bibinfo{pages}{2025--2047}
  (\bibinfo{year}{2016}).
\newblock \urlprefix\url{https://doi.org/10.1088/0031-9155/61/5/2025}.

\bibitem{jpet}
\bibinfo{author}{Moskal, P.} \& \bibinfo{author}{Stepien, E.}
\newblock \bibinfo{title}{Prospects and clinical perspectives of total-body pet
  imaging using plastic scintillators}.
\newblock \emph{\bibinfo{journal}{PET Clinics}} \textbf{\bibinfo{volume}{15}}
  (\bibinfo{year}{2020}).
\newblock \urlprefix\url{https://doi.org/10.1016/j.cpet.2020.06.009}.

\bibitem{3photonDec}
\bibinfo{author}{Hiesmayr, B.~C.} \& \bibinfo{author}{Moskal, P.}
\newblock \bibinfo{title}{Genuine multipartite entanglement in the 3-photon
  decay of positronium}.
\newblock \emph{\bibinfo{journal}{Scientific Reports}}
  \textbf{\bibinfo{volume}{7}}, \bibinfo{pages}{15349} (\bibinfo{year}{2017}).
\newblock \urlprefix\url{https://doi.org/10.1038/s41598-017-15356-y}.

\bibitem{comptonMUB}
\bibinfo{author}{Hiesmayr, B.~C.} \& \bibinfo{author}{Moskal, P.}
\newblock \bibinfo{title}{Witnessing entanglement in compton scattering
  processes via mutually unbiased bases}.
\newblock \emph{\bibinfo{journal}{Scientific Reports}}
  \textbf{\bibinfo{volume}{9}}, \bibinfo{pages}{8166} (\bibinfo{year}{2019}).
\newblock \urlprefix\url{https://doi.org/10.1038/s41598-019-44570-z}.

\bibitem{highDimQCReview}
\bibinfo{author}{Cozzolino, D.}, \bibinfo{author}{Da~Lio, B.},
  \bibinfo{author}{Bacco, D.} \& \bibinfo{author}{Oxenløwe, L.~K.}
\newblock \bibinfo{title}{High-dimensional quantum communication: Benefits,
  progress, and future challenges}.
\newblock \emph{\bibinfo{journal}{Advanced Quantum Technologies}}
  \textbf{\bibinfo{volume}{2}}, \bibinfo{pages}{1900038}
  (\bibinfo{year}{2019}).
\newblock \urlprefix\url{https://doi.org/10.1002/qute.201900038}.

\bibitem{qditsQCWang}
\bibinfo{author}{Wang, Y.}, \bibinfo{author}{Hu, Z.}, \bibinfo{author}{Sanders,
  B.~C.} \& \bibinfo{author}{Kais, S.}
\newblock \bibinfo{title}{Qudits and high-dimensional quantum computing}.
\newblock \emph{\bibinfo{journal}{Frontiers in Physics}}
  \textbf{\bibinfo{volume}{8}}, \bibinfo{pages}{479} (\bibinfo{year}{2020}).
\newblock
  \urlprefix\url{https://www.frontiersin.org/article/10.3389/fphy.2020.589504}.

\bibitem{bellStates}
\bibinfo{author}{Braunstein, S.~L.}, \bibinfo{author}{Mann, A.} \&
  \bibinfo{author}{Revzen, M.}
\newblock \bibinfo{title}{Maximal violation of bell inequalities for mixed
  states}.
\newblock \emph{\bibinfo{journal}{Phys. Rev. Lett.}}
  \textbf{\bibinfo{volume}{68}}, \bibinfo{pages}{3259--3261}
  (\bibinfo{year}{1992}).
\newblock \urlprefix\url{https://link.aps.org/doi/10.1103/PhysRevLett.68.3259}.

\bibitem{Sych}
\bibinfo{author}{Sych, D.} \& \bibinfo{author}{Leuchs, G.}
\newblock \bibinfo{title}{A complete basis of generalized bell states}.
\newblock \emph{\bibinfo{journal}{New Journal of Physics}}
  \textbf{\bibinfo{volume}{11}}, \bibinfo{pages}{013006}
  (\bibinfo{year}{2009}).
\newblock \urlprefix\url{https://doi.org/10.1088/1367-2630/11/1/013006}.

\bibitem{teleportingWeyl}
\bibinfo{author}{Bennett, C.~H.} \emph{et~al.}
\newblock \bibinfo{title}{Teleporting an unknown quantum state via dual
  classical and einstein-podolsky-rosen channels}.
\newblock \emph{\bibinfo{journal}{Phys. Rev. Lett.}}
  \textbf{\bibinfo{volume}{70}}, \bibinfo{pages}{1895--1899}
  (\bibinfo{year}{1993}).
\newblock \urlprefix\url{https://link.aps.org/doi/10.1103/PhysRevLett.70.1895}.

\bibitem{baumgartner1}
\bibinfo{author}{Baumgartner, B.}, \bibinfo{author}{Hiesmayr, B.~C.} \&
  \bibinfo{author}{Narnhofer, H.}
\newblock \bibinfo{title}{A special simplex in the state space for entangled
  qudits}.
\newblock \emph{\bibinfo{journal}{J. Phys. A: Math. Theor.}}
  \textbf{\bibinfo{volume}{40}}, \bibinfo{pages}{7919} (\bibinfo{year}{2007}).
\newblock \urlprefix\url{https://doi.org/10.1088/1751-8113/40/28/S03}.

\bibitem{peres}
\bibinfo{author}{Peres, A.}
\newblock \bibinfo{title}{Separability criterion for density matrices}.
\newblock \emph{\bibinfo{journal}{Phys. Rev. Lett.}}
  \textbf{\bibinfo{volume}{77}}, \bibinfo{pages}{1413--1415}
  (\bibinfo{year}{1996}).
\newblock \urlprefix\url{https://link.aps.org/doi/10.1103/PhysRevLett.77.1413}.

\bibitem{horodeckiCriterion}
\bibinfo{author}{Horodecki, M.}, \bibinfo{author}{Horodecki, P.} \&
  \bibinfo{author}{Horodecki, R.}
\newblock \bibinfo{title}{Separability of mixed states: necessary and
  sufficient conditions}.
\newblock \emph{\bibinfo{journal}{Physics Letters A}}
  \textbf{\bibinfo{volume}{223}}, \bibinfo{pages}{1--8} (\bibinfo{year}{1996}).
\newblock
  \urlprefix\url{https://www.sciencedirect.com/science/article/pii/S0375960196007062}.

\bibitem{distillation}
\bibinfo{author}{Horodecki, M.}, \bibinfo{author}{Horodecki, P.} \&
  \bibinfo{author}{Horodecki, R.}
\newblock \bibinfo{title}{Mixed-state entanglement and distillation: Is there a
  ``bound'' entanglement in nature?}
\newblock \emph{\bibinfo{journal}{Phys. Rev. Lett.}}
  \textbf{\bibinfo{volume}{80}}, \bibinfo{pages}{5239--5242}
  (\bibinfo{year}{1998}).
\newblock \urlprefix\url{https://link.aps.org/doi/10.1103/PhysRevLett.80.5239}.

\bibitem{Halder_BE}
\bibinfo{author}{Bej, P.} \& \bibinfo{author}{Halder, S.}
\newblock \bibinfo{title}{Unextendible product bases, bound entangled states,
  and the range criterion}.
\newblock \emph{\bibinfo{journal}{Physics Letters A}}
  \textbf{\bibinfo{volume}{386}}, \bibinfo{pages}{126992}
  (\bibinfo{year}{2021}).
\newblock
  \urlprefix\url{https://www.sciencedirect.com/science/article/pii/S0375960120308598}.

\bibitem{lockhart_e}
\bibinfo{author}{Lockhart, J.}, \bibinfo{author}{G\"uhne, O.} \&
  \bibinfo{author}{Severini, S.}
\newblock \bibinfo{title}{Entanglement properties of quantum grid states}.
\newblock \emph{\bibinfo{journal}{Phys. Rev. A}} \textbf{\bibinfo{volume}{97}},
  \bibinfo{pages}{062340} (\bibinfo{year}{2018}).
\newblock \urlprefix\url{https://link.aps.org/doi/10.1103/PhysRevA.97.062340}.

\bibitem{Bruss_BE}
\bibinfo{author}{Bru\ss{}, D.} \& \bibinfo{author}{Peres, A.}
\newblock \bibinfo{title}{Construction of quantum states with bound
  entanglement}.
\newblock \emph{\bibinfo{journal}{Phys. Rev. A}} \textbf{\bibinfo{volume}{61}},
  \bibinfo{pages}{030301} (\bibinfo{year}{2000}).
\newblock \urlprefix\url{https://link.aps.org/doi/10.1103/PhysRevA.61.030301}.

\bibitem{Slater2019JaggedIO}
\bibinfo{author}{Slater, P.~B.}
\newblock \bibinfo{title}{Jagged islands of bound entanglement and
  witness-parameterized probabilities.}
\newblock \emph{\bibinfo{journal}{arXiv: Quantum Physics}}
  (\bibinfo{year}{2019}).
\newblock \urlprefix\url{https://doi.org/10.48550/arXiv.1905.09228}.

\bibitem{choi}
\bibinfo{author}{Choi, M.-D.}
\newblock \bibinfo{title}{Some assorted inequalities for positive linear maps
  on c*-algebras}.
\newblock \emph{\bibinfo{journal}{Journal of Operator Theory}}
  \textbf{\bibinfo{volume}{4}} (\bibinfo{year}{1980}).

\bibitem{Chru_EW}
\bibinfo{author}{Chru{\'{s}}ci{\'{n}}ski, D.} \& \bibinfo{author}{Sarbicki, G.}
\newblock \bibinfo{title}{Entanglement witnesses: construction, analysis and
  classification}.
\newblock \emph{\bibinfo{journal}{J. Phys. A: Math. Theor.}}
  \textbf{\bibinfo{volume}{47}}, \bibinfo{pages}{483001}
  (\bibinfo{year}{2014}).
\newblock \urlprefix\url{https://doi.org/10.1088/1751-8113/47/48/483001}.

\bibitem{BaeLinkingED}
\bibinfo{author}{Kalev, A.} \& \bibinfo{author}{Bae, J.}
\newblock \bibinfo{title}{Optimal approximate transpose map via quantum designs
  and its applications to entanglement detection}.
\newblock \emph{\bibinfo{journal}{Phys. Rev. A}} \textbf{\bibinfo{volume}{87}},
  \bibinfo{pages}{062314} (\bibinfo{year}{2013}).
\newblock \urlprefix\url{https://link.aps.org/doi/10.1103/PhysRevA.87.062314}.

\bibitem{Bae_structPhysApprox}
\bibinfo{author}{Bae, J.}
\newblock \bibinfo{title}{Designing quantum information processing via
  structural physical approximation}.
\newblock \emph{\bibinfo{journal}{Reports on Progress in Physics}}
  \textbf{\bibinfo{volume}{80}}, \bibinfo{pages}{104001}
  (\bibinfo{year}{2017}).
\newblock \urlprefix\url{https://doi.org/10.1088/1361-6633/aa7d45}.

\bibitem{Korbicz2008StructuralAT}
\bibinfo{author}{Korbicz, J.~K.}, \bibinfo{author}{Almeida, M.~L.},
  \bibinfo{author}{Bae, J.}, \bibinfo{author}{Lewenstein, M.} \&
  \bibinfo{author}{Ac\'{\i}n, A.}
\newblock \bibinfo{title}{Structural approximations to positive maps and
  entanglement-breaking channels}.
\newblock \emph{\bibinfo{journal}{Phys. Rev. A}} \textbf{\bibinfo{volume}{78}},
  \bibinfo{pages}{062105} (\bibinfo{year}{2008}).
\newblock \urlprefix\url{https://link.aps.org/doi/10.1103/PhysRevA.78.062105}.

\bibitem{Huber_hdent}
\bibinfo{author}{Huber, M.}, \bibinfo{author}{Mintert, F.},
  \bibinfo{author}{Gabriel, A.} \& \bibinfo{author}{Hiesmayr, B.~C.}
\newblock \bibinfo{title}{Detection of high-dimensional genuine multipartite
  entanglement of mixed states}.
\newblock \emph{\bibinfo{journal}{Phys. Rev. Lett.}}
  \textbf{\bibinfo{volume}{104}}, \bibinfo{pages}{210501}
  (\bibinfo{year}{2010}).
\newblock
  \urlprefix\url{https://link.aps.org/doi/10.1103/PhysRevLett.104.210501}.

\bibitem{augus_optEw}
\bibinfo{author}{Augusiak, R.}, \bibinfo{author}{Bae, J.},
  \bibinfo{author}{Tura~Brugués, J.} \& \bibinfo{author}{Lewenstein, M.}
\newblock \bibinfo{title}{Checking the optimality of entanglement witnesses: An
  application to structural physical approximations}.
\newblock \emph{\bibinfo{journal}{J. Phys. A: Math. Theor.}}
  \textbf{\bibinfo{volume}{47}} (\bibinfo{year}{2014}).
\newblock
  \urlprefix\url{https://iopscience.iop.org/article/10.1088/1751-8113/47/6/065301}.

\bibitem{hiesmayrLoeffler}
\bibinfo{author}{Hiesmayr, B.~C.} \& \bibinfo{author}{Löffler, W.}
\newblock \bibinfo{title}{Complementarity reveals bound entanglement of two
  twisted photons}.
\newblock \emph{\bibinfo{journal}{New J. Phys}} \textbf{\bibinfo{volume}{15}},
  \bibinfo{pages}{083036} (\bibinfo{year}{2013}).
\newblock \urlprefix\url{https://doi.org/10.1088/1367-2630/15/8/083036}.

\bibitem{nphard}
\bibinfo{author}{Gurvits, L.}
\newblock \bibinfo{title}{Classical deterministic complexity of edmonds'
  problem and quantum entanglement}.
\newblock In \emph{\bibinfo{booktitle}{Proceedings of the Thirty-Fifth Annual
  ACM Symposium on Theory of Computing}}, STOC '03, \bibinfo{pages}{10–19}
  (\bibinfo{publisher}{Association for Computing Machinery},
  \bibinfo{address}{New York, NY, USA}, \bibinfo{year}{2003}).
\newblock \urlprefix\url{https://doi.org/10.1145/780542.780545}.

\bibitem{nphard-strong}
\bibinfo{author}{Gharibian, S.}
\newblock \bibinfo{title}{Strong np-hardness of the quantum separability
  problem}.
\newblock \emph{\bibinfo{journal}{Quantum Information and Computation}}
  \textbf{\bibinfo{volume}{10}} (\bibinfo{year}{2008}).
\newblock \urlprefix\url{https://doi.org/10.26421/QIC10.3-4-11}.

\bibitem{Werner_2001}
\bibinfo{author}{Werner, R.~F.}
\newblock \bibinfo{title}{All teleportation and dense coding schemes}.
\newblock \emph{\bibinfo{journal}{Journal of Physics A: Mathematical and
  General}} \textbf{\bibinfo{volume}{34}}, \bibinfo{pages}{7081--7094}
  (\bibinfo{year}{2001}).
\newblock \urlprefix\url{https://doi.org/10.1088/0305-4470/34/35/332}.

\bibitem{baumgartnerHiesmayr}
\bibinfo{author}{Baumgartner, B.}, \bibinfo{author}{Hiesmayr, B.~C.} \&
  \bibinfo{author}{Narnhofer, H.}
\newblock \bibinfo{title}{State space for two qutrits has a phase space
  structure in its core}.
\newblock \emph{\bibinfo{journal}{Phys. Rev. A}} \textbf{\bibinfo{volume}{74}},
  \bibinfo{pages}{032327} (\bibinfo{year}{2006}).
\newblock \urlprefix\url{https://link.aps.org/doi/10.1103/PhysRevA.74.032327}.

\bibitem{EWs}
\bibinfo{author}{Terhal, B.~M.}
\newblock \bibinfo{title}{Bell inequalities and the separability criterion}.
\newblock \emph{\bibinfo{journal}{Physics Letters A}}
  \textbf{\bibinfo{volume}{271}}, \bibinfo{pages}{319--326}
  (\bibinfo{year}{2000}).
\newblock
  \urlprefix\url{https://www.sciencedirect.com/science/article/pii/S0375960100004011}.

\bibitem{EW20}
\bibinfo{author}{Bae, J.}, \bibinfo{author}{Chruściński, D.} \&
  \bibinfo{author}{Hiesmayr, B.~C.}
\newblock \bibinfo{title}{Mirrored entanglement witnesses}.
\newblock \emph{\bibinfo{journal}{npj Quantum Information}}
  \textbf{\bibinfo{volume}{6}} (\bibinfo{year}{2020}).
\newblock \urlprefix\url{https://doi.org/10.1038/s41534-020-0242-z}.

\bibitem{PoppACS}
\bibinfo{author}{Popp, C.} \& \bibinfo{author}{Hiesmayr, B.~C.}
\newblock \bibinfo{title}{Almost complete solution for the np-hard separability
  problem of bell diagonal qutrits}.
\newblock \emph{\bibinfo{journal}{Scientific Reports}}
  \textbf{\bibinfo{volume}{12}}, \bibinfo{pages}{12472} (\bibinfo{year}{2022}).
\newblock \urlprefix\url{https://doi.org/10.1038/s41598-022-16225-z}.

\bibitem{volOfSep1}
\bibinfo{author}{\ifmmode~\dot{Z}\else \.{Z}\fi{}yczkowski, K.},
  \bibinfo{author}{Horodecki, P.}, \bibinfo{author}{Sanpera, A.} \&
  \bibinfo{author}{Lewenstein, M.}
\newblock \bibinfo{title}{Volume of the set of separable states}.
\newblock \emph{\bibinfo{journal}{Phys. Rev. A}} \textbf{\bibinfo{volume}{58}},
  \bibinfo{pages}{883--892} (\bibinfo{year}{1998}).
\newblock \urlprefix\url{https://link.aps.org/doi/10.1103/PhysRevA.58.883}.

\bibitem{volOfSep2}
\bibinfo{author}{Zyczkowski, K.}
\newblock \bibinfo{title}{Volume of the set of separable states. ii}.
\newblock \emph{\bibinfo{journal}{Phys. Rev. A}} \textbf{\bibinfo{volume}{60}},
  \bibinfo{pages}{3496--3507} (\bibinfo{year}{1999}).
\newblock \urlprefix\url{https://link.aps.org/doi/10.1103/PhysRevA.60.3496}.

\bibitem{magicBellBasis}
\bibinfo{author}{Hill, S.~A.} \& \bibinfo{author}{Wootters, W.~K.}
\newblock \bibinfo{title}{Entanglement of a pair of quantum bits}.
\newblock \emph{\bibinfo{journal}{Phys. Rev. Lett.}}
  \textbf{\bibinfo{volume}{78}}, \bibinfo{pages}{5022--5025}
  (\bibinfo{year}{1997}).
\newblock \urlprefix\url{https://link.aps.org/doi/10.1103/PhysRevLett.78.5022}.

\bibitem{baumgartnerHiesmayr2}
\bibinfo{author}{Baumgartner, B.}, \bibinfo{author}{Hiesmayr, B.~C.} \&
  \bibinfo{author}{Narnhofer, H.}
\newblock \bibinfo{title}{The geometry of bipartite qutrits including bound
  entanglement}.
\newblock \emph{\bibinfo{journal}{Physics Letters A}}
  \textbf{\bibinfo{volume}{372}}, \bibinfo{pages}{2190--2195}
  (\bibinfo{year}{2008}).
\newblock
  \urlprefix\url{https://www.sciencedirect.com/science/article/pii/S0375960107016507}.

\bibitem{plotlyJs}
\bibinfo{title}{Plotly graphing libraries (v0.18.6)}.
\newblock \urlprefix\url{https://plotly.com/julia/}.

\bibitem{BaeQuasipure}
\bibinfo{author}{Bae, J.} \emph{et~al.}
\newblock \bibinfo{title}{Detection and typicality of bound entangled states}.
\newblock \emph{\bibinfo{journal}{Phys. Rev. A}} \textbf{\bibinfo{volume}{80}},
  \bibinfo{pages}{022317} (\bibinfo{year}{2009}).
\newblock \urlprefix\url{https://link.aps.org/doi/10.1103/PhysRevA.80.022317}.

\bibitem{simplexSampling}
\bibinfo{author}{Willms, A.~R.}
\newblock \bibinfo{title}{{Uniform Sampling on the Standard Simplex}}.
\newblock \emph{\bibinfo{journal}{Missouri Journal of Mathematical Sciences}}
  \textbf{\bibinfo{volume}{33}}, \bibinfo{pages}{119--124}
  (\bibinfo{year}{2021}).
\newblock \urlprefix\url{https://doi.org/10.35834/2021/3301119}.

\bibitem{realignment}
\bibinfo{author}{Chen, K.} \& \bibinfo{author}{Wu, L.-A.}
\newblock \bibinfo{title}{A matrix realignment method for recognizing
  entanglement}.
\newblock \emph{\bibinfo{journal}{Quantum Information and Computation}}
  \textbf{\bibinfo{volume}{3}} (\bibinfo{year}{2002}).
\newblock \urlprefix\url{https://doi.org/10.48550/arXiv.quant-ph/0205017}.

\bibitem{woottersConcurrence}
\bibinfo{author}{Wootters, W.~K.}
\newblock \bibinfo{title}{Entanglement of formation of an arbitrary state of
  two qubits}.
\newblock \emph{\bibinfo{journal}{Phys. Rev. Lett.}}
  \textbf{\bibinfo{volume}{80}}, \bibinfo{pages}{2245--2248}
  (\bibinfo{year}{1998}).
\newblock \urlprefix\url{https://link.aps.org/doi/10.1103/PhysRevLett.80.2245}.

\bibitem{woottersMUB}
\bibinfo{author}{Wootters, W.~K.} \& \bibinfo{author}{Fields, B.~D.}
\newblock \bibinfo{title}{Optimal state-determination by mutually unbiased
  measurements}.
\newblock \emph{\bibinfo{journal}{Annals of Physics}}
  \textbf{\bibinfo{volume}{191}}, \bibinfo{pages}{363--381}
  (\bibinfo{year}{1989}).
\newblock
  \urlprefix\url{https://www.sciencedirect.com/science/article/pii/0003491689903229}.

\bibitem{BandyopadhyayMUB}
\bibinfo{author}{Bandyopadhyay, S.}, \bibinfo{author}{Boykin, P.~O.},
  \bibinfo{author}{Roychowdhury, V.~P.} \& \bibinfo{author}{Vatan, F.}
\newblock \bibinfo{title}{A new proof for the existence of mutually unbiased
  bases}.
\newblock \emph{\bibinfo{journal}{Algorithmica}} \textbf{\bibinfo{volume}{34}},
  \bibinfo{pages}{512--528} (\bibinfo{year}{2002}).
\newblock \urlprefix\url{https://doi.org/10.1007/s00453-002-0980-7}.

\bibitem{bae2021measurements}
\bibinfo{author}{Bae, J.}, \bibinfo{author}{Bera, A.},
  \bibinfo{author}{Chruściński, D.}, \bibinfo{author}{Hiesmayr, B.~C.} \&
  \bibinfo{author}{McNulty, D.}
\newblock \bibinfo{title}{How many measurements are needed to detect bound
  entangled states?} (\bibinfo{year}{2021}).
\newblock \eprint{2108.01109}.

\bibitem{SpenglerMUB}
\bibinfo{author}{Spengler, C.}, \bibinfo{author}{Huber, M.},
  \bibinfo{author}{Brierley, S.}, \bibinfo{author}{Adaktylos, T.} \&
  \bibinfo{author}{Hiesmayr, B.~C.}
\newblock \bibinfo{title}{Entanglement detection via mutually unbiased bases}.
\newblock \emph{\bibinfo{journal}{Phys. Rev. A}} \textbf{\bibinfo{volume}{86}},
  \bibinfo{pages}{022311} (\bibinfo{year}{2012}).
\newblock \urlprefix\url{https://link.aps.org/doi/10.1103/PhysRevA.86.022311}.

\bibitem{ineqMUBs}
\bibinfo{author}{Hiesmayr, B.~C.} \emph{et~al.}
\newblock \bibinfo{title}{Detecting entanglement can be more effective with
  inequivalent mutually unbiased bases}.
\newblock \emph{\bibinfo{journal}{New Journal of Physics}}
  \textbf{\bibinfo{volume}{23}}, \bibinfo{pages}{093018}
  (\bibinfo{year}{2021}).
\newblock \urlprefix\url{https://doi.org/10.1088/1367-2630/ac20ea}.

\bibitem{spenglerHuberHiesmayr2}
\bibinfo{author}{Spengler, C.}, \bibinfo{author}{Huber, M.} \&
  \bibinfo{author}{Hiesmayr, B.~C.}
\newblock \bibinfo{title}{A composite parameterization of unitary groups,
  density matrices and subspaces}.
\newblock \emph{\bibinfo{journal}{J. Phys. A: Math. Theor}}
  \textbf{\bibinfo{volume}{43}}, \bibinfo{pages}{385306}
  (\bibinfo{year}{2010}).
\newblock \urlprefix\url{https://doi.org/10.1088/1751-8113/43/38/385306}.

\bibitem{lazysets}
\bibinfo{author}{Forets, M.} \& \bibinfo{author}{Schilling, C.}
\newblock \bibinfo{title}{Lazysets.jl: Scalable symbolic-numeric set
  computations$^*$}.
\newblock \emph{\bibinfo{journal}{Proceedings of the JuliaCon Conferences}}
  \textbf{\bibinfo{volume}{1}}, \bibinfo{pages}{97} (\bibinfo{year}{2021}).
\newblock \urlprefix\url{https://doi.org/10.21105/jcon.00097}.

\bibitem{weylRepCrit}
\bibinfo{author}{Chru{\'{s}}ci{\'{n}}ski, D.} \& \bibinfo{author}{Pittenger,
  A.~O.}
\newblock \bibinfo{title}{Generalized circulant densities and a sufficient
  condition for separability}.
\newblock \emph{\bibinfo{journal}{J. Phys. A: Math. Theor.}}
  \textbf{\bibinfo{volume}{41}}, \bibinfo{pages}{385301}
  (\bibinfo{year}{2008}).
\newblock \urlprefix\url{https://doi.org/10.1088/1751-8113/41/38/385301}.

\bibitem{geoPic}
\bibinfo{author}{Bertlmann, R.~A.}, \bibinfo{author}{Narnhofer, H.} \&
  \bibinfo{author}{Thirring, W.}
\newblock \bibinfo{title}{Geometric picture of entanglement and bell
  inequalities}.
\newblock \emph{\bibinfo{journal}{Phys. Rev. A}} \textbf{\bibinfo{volume}{66}},
  \bibinfo{pages}{032319} (\bibinfo{year}{2002}).
\newblock \urlprefix\url{https://link.aps.org/doi/10.1103/PhysRevA.66.032319}.

\bibitem{geoQS}
\bibinfo{author}{Bengtsson, I.} \& \bibinfo{author}{{\.Z}yczkowski, K.}
\newblock \emph{\bibinfo{title}{Geometry of Quantum States: An Introduction to
  Quantum Entanglement}} (\bibinfo{publisher}{Cambridge University Press},
  \bibinfo{year}{2006}).

\bibitem{SCMCS}
\bibinfo{author}{Li, W.}, \bibinfo{author}{Han, R.}, \bibinfo{author}{Shang,
  J.}, \bibinfo{author}{Ng, H.~K.} \& \bibinfo{author}{Englert, B.-G.}
\newblock \bibinfo{title}{Sequentially constrained monte carlo sampler for
  quantum states} (\bibinfo{year}{2021}).
\newblock \urlprefix\url{https://arxiv.org/abs/2109.14215}.

\bibitem{BellDiagonalQudits}
\bibinfo{author}{Popp, C.}
\newblock \bibinfo{title}{Belldiagonalqudits: A package for entanglement
  analyses of mixed maximally entangled qudits}.
\newblock \emph{\bibinfo{journal}{Journal of Open Source Software}}
  \textbf{\bibinfo{volume}{8}}, \bibinfo{pages}{4924} (\bibinfo{year}{2023}).
\newblock \urlprefix\url{https://doi.org/10.21105/joss.04924}.

\end{thebibliography}

\section*{Acknowledgments}
B.C.H. and C.P. acknowledge gratefully that this research was funded in whole, or in
part, by the  Austrian Science Fund (FWF) project P36102-N. For the purpose of
open access, the author has applied a CC BY public copyright licence to any
Author Accepted Manuscript version arising from this submission.

\section*{Author contributions statement}
C.P. implemented the software, developed the new methods and analyzed the data.\\
B.C.H. revised the analyses and results and proposed improvements. \\
C.P and B.C.H. both edited the manuscript.

\section*{Additional information}
Correspondence and requests for materials should be addressed to C.P..

\section*{Appendix}
\label{sec:appendix}

\subsection*{A1: MUB sets for $d=3$ and $d=4$}
\label{sec:mub-sets}
The following sets of bases are known to be mutually unbiased \cite{SpenglerMUB}
and non-decomposable, i.e. able to detect bound entanglement
\cite{bae2021measurements}. Note, that other MUBs exist, which generally detect
different sets of entangled states \cite{bae2021measurements, ineqMUBs}. For $d=3$ and $d=4$ and $w = e^{\frac{2 \pi i}{d}}$,
the used bases with basis vectors combined to matrix columns and represented in the
computational basis read as follows:
\\ \ \\
\textbf{d = 3}:
\begin{gather*}
  B_1 =  
\begin{pmatrix}
    1 & 0 & 0\\
    0 & 1 & 0\\
    0 & 0 & 1
  \end{pmatrix},  
~B_2 =  \frac{1}{\sqrt{3}}
\begin{pmatrix}
    1 & 1 & 1\\
    1 & w & w^2\\
    1 & w^2 & w
  \end{pmatrix},
~B_3 =  \frac{1}{\sqrt{3}}
\begin{pmatrix}
    1 & 1 & 1\\
    w & w^2 & 1\\
    w & 1 & w^2
  \end{pmatrix},
~B_4 =  \frac{1}{\sqrt{3}}
\begin{pmatrix}
    1 & 1 & 1\\
    w^2 & 1 & w\\
    w^2 & w & 1
  \end{pmatrix}
\end{gather*}
\textbf{d = 4}:
 \begin{gather*}
   B_1 = \begin{pmatrix}
    1 & 0 & 0 & 0\\
    0 & 1 & 0 & 0\\
    0 & 0 & 1 & 0\\
    0 & 0 & 0 & 1
  \end{pmatrix},
  ~B_2 = \frac{1}{2} \begin{pmatrix}
    1 & 1 & 1 & 1\\
    1 & 1 & -1 & -1\\
    1 & -1 & -1 & 1\\
    1 & -1 & 1 & -1
  \end{pmatrix},
  ~B_3 = \frac{1}{2} \begin{pmatrix}
    1 & 1 & 1 & 1\\
    -i & -i & i & i\\
    -i & i & i & -i\\
    -1 & 1 & -1 & 1
  \end{pmatrix}, \\ 
  ~B_4 = \frac{1}{2} \begin{pmatrix}
    1 & 1 & 1 & 1\\
    -1 & -1 & 1 & 1\\
    -i & i & i & -i\\
    -i & i & -i & i
  \end{pmatrix},
  ~B_5 = \frac{1}{2} \begin{pmatrix}
    1 & 1 & 1 & 1\\
    -i & -i & i & i\\
    -1 & 1 & 1 & -1\\
    -i & i & -i & i
  \end{pmatrix}
\end{gather*}

\end{document}